\documentclass[11pt,a4paper]{article}
\usepackage[hyperref]{naaclhlt2019}
\usepackage{times}
\usepackage{latexsym}
\usepackage{url}
\usepackage{amsfonts}
\usepackage{arydshln}
\usepackage{booktabs}
\usepackage{multirow}
\usepackage{helvet}
\usepackage{color}
\usepackage{courier}
\usepackage[utf8]{inputenc}
\usepackage{graphicx}
\usepackage{todonotes}
\usepackage{enumitem}
\hypersetup{draft}

\newcommand{\Ni}{({\em i})~}
\newcommand{\Nii}{({\em ii})~}
\newcommand{\Niii}{({\em iii})~}

\aclfinalcopy 
\def\aclpaperid{590} 

\title{Multi-Task Ordinal Regression for Jointly Predicting\\the Trustworthiness and the Leading Political Ideology of News Media}

\author{%
	Ramy Baly$^1$,
	Georgi Karadzhov$^2$,
	Abdelrhman Saleh$^3$,\\
	\bf James Glass$^1$,
	Preslav Nakov$^4$\\
    $^1$MIT Computer Science and Artificial Intelligence Laboratory, MA, USA\\
    $^2$SiteGround Hosting EOOD, Bulgaria,
	$^3$Harvard University, MA, USA\\
    $^4$Qatar Computing Research Institute, HBKU, Qatar\\
    {\tt \{baly, glass\}@mit.edu},
    {\tt georgi.m.karadjov@gmail.com}\\
    {\tt abdelrhman\_saleh@college.harvard.edu},
    {\tt pnakov@hbku.edu.qa}
 }

\date{}

\begin{document}
\maketitle

\begin{abstract}
In the context of fake news, bias, and propaganda, we study two important but relatively under-explored problems: \Ni \emph{trustworthiness estimation} (on a 3-point scale) and \Nii \emph{political ideology detection} (left/right bias on a 7-point scale) of entire news outlets, as opposed to evaluating individual articles.
In particular, we propose a multi-task ordinal regression framework that models the two problems jointly.
This is motivated by the observation that hyper-partisanship is often linked to low trustworthiness, e.g.,~appealing to emotions rather than sticking to the facts, 
while center media tend to be generally more impartial and trustworthy.
We further use several auxiliary tasks, modeling centrality, hyper-partisanship, as well as left-vs.-right bias on a coarse-grained scale.
The evaluation results show sizable performance gains by the joint models over models that target the problems in isolation.
\end{abstract}

\section{Introduction}\label{sec:intro}

Recent years have seen the rise of social media, which has enabled people to virtually share information with a large number of users without regulation or quality control.
On the bright side, this has given an opportunity for anyone to become a content creator, and has also enabled a much faster information dissemination.
However, it has also opened the door for malicious users to spread disinformation and misinformation much faster, enabling them to easily reach audience at a scale that was never possible before.
In some cases, this involved building sophisticated profiles for individuals based on a combination of psychological characteristics, meta-data, demographics, and location, and then micro-targeting them with personalized ``fake news'' with the aim of achieving some political or financial gains \cite{Lazer1094,Vosoughi1146}.

\noindent 
A number of fact-checking initiatives have been launched so far, both manual and automatic, but the whole enterprise remains in a state of crisis: by the time a claim is finally fact-checked, it could have reached millions of users, and the harm caused could hardly be undone. An arguably more promising direction is to focus on fact-checking entire news outlets, which can be done in advance. Then, we could fact-check the news before they were even written: by checking how trustworthy the outlets that published them are.
Knowing the reliability of a medium is important not only when fact-checking a claim \cite{Popat:2017:TLE:3041021.3055133,DBLP:conf/aaai/NguyenKLW18}, but also when solving article-level tasks such as ``fake news'' and click-bait detection \cite{brill2001online,finberg2002digital,Hardalov2016,RANLP2017:clickbait,desarkar-yang-mukherjee:2018:C18-1,Pan:KG:2018,prezrosas-EtAl:2018:C18-1}

\begin{figure}[tbh]
    \centering
    \includegraphics[width=0.43\textwidth]{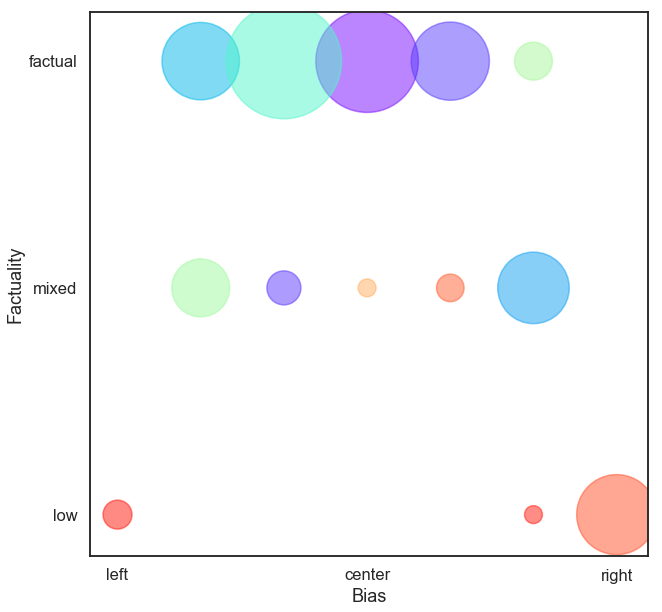}
    \caption{Correlation between bias and factuality for the news outlets in the Media Bias/Fact Check website.}
    \label{fig:mapping}
\end{figure}

Political ideology (or left/right bias) is a related characteristic, e.g., extreme left/right media tend to be propagandistic, while center media are more factual, and thus generally more trustworthy.
This connection can be clearly seen in Figure~\ref{fig:mapping}.

\noindent Despite the connection between factuality and bias, previous research has addressed them as independent tasks, even when the underlying dataset had annotations for both \cite{baly2018predicting}.
In contrast, here we solve them jointly.
Our contributions can be summarized as follows:

\begin{itemize}[noitemsep,topsep=0pt]
  \item We study an under-explored but arguably important problem: predicting the factuality of reporting of news media.
  Moreover, unlike previous work, we do this jointly with the task of predicting political bias.
  \item As factuality and bias are naturally defined on an ordinal scale (factuality: from \emph{low} to \emph{high}, and bias: from \emph{extreme-left} to \emph{extreme-right}), we address them as ordinal regression. Using multi-task ordinal regression is novel for these tasks, and it is also an under-explored direction in machine learning in general.
  \item We design a variety of auxiliary subtasks from the bias labels: modeling centrality, hyper-partisanship, as well as left-vs.-right bias on a coarse-grained scale.

\end{itemize}

\section{Related Work}
\label{sec:related_work}

\paragraph{Factuality of Reporting}

Previous work has modeled the factuality of reporting at the medium level by checking the general stance of the target medium with respect to known manually fact-checked claims, without access to gold labels about the overall medium-level factuality of reporting \cite{mukherjee2015leveraging,popat2016credibility,Popat:2017:TLE:3041021.3055133,Popat:2018:CCL:3184558.3186967}. 

The trustworthiness of Web sources has also been studied from a Data Analytics perspective, e.g., \citet{Dong:2015:KTE:2777598.2777603} proposed that a trustworthy source is one that contains very few false claims.
In social media, there has been research targeting the user, e.g., finding malicious users \cite{mihaylov-nakov:2016:P16-2,AAAI2018:factchecking,InternetResearchJournal:2018}, \emph{sockpuppets} \cite{Maity:2017:DSS:3022198.3026360}, \emph{Internet water army} \cite{Chen:2013:BIW:2492517.2492637}, and \emph{seminar users} \cite{SeminarUsers2017}. 

Unlike the above work, here we study source reliability as a task in its own right, using manual gold annotations specific for the task and assigned by independent fact-checking journalists. Moreover, we address the problem as one of ordinal regression on a three-point scale, and we solve it jointly with political ideology prediction in a multi-task learning setup, using several auxiliary tasks.

\paragraph{Predicting Political Ideology}

In previous work, political ideology, also known as media bias, was used as a feature for ``fake news'' detection \cite{DBLP:journals/corr/abs-1803-10124}.
It has also been the target of classification, e.g.,~\citet{Horne:2018:ANL:3184558.3186987} predicted whether an article is biased 
(\emph{political} or \emph{bias}) 
vs. unbiased. 
Similarly, \citet{DBLP:journals/corr/PotthastKRBS17} classified the bias in a target article as
(\emph{i})~left vs. right vs. mainstream, or as
(\emph{ii})~hyper-partisan vs. mainstream. 
Left-vs-right bias classification at the article level was also explored by \citet{kulkarni2018multi}, who modeled both the textual and the URL contents of the target article. 
There has been also work targeting bias at the phrase or the sentence level \cite{P14-1105}, focusing on political speeches \cite{D13-1010} or legislative documents \cite{Gerrish:2011:PLR:3104482.3104544}, or targeting users in Twitter \cite{P17-1068}.
Another line of related work focuses on propaganda, which can be seen as a form of extreme bias \cite{rashkin-EtAl:2017:EMNLP2017,AAAI2019:proppy,Barron:19}. 
See also a recent position paper \cite{Pitoura:2018:MBO:3186549.3186553} and an overview paper on bias on the Web \cite{Baeza-Yates:2018:BW:3229066.3209581}. 
Unlike the above work, here we focus on predicting the political ideology of news media outlets. 

In our previous work \cite{baly2018predicting}, we did target the political bias of entire news outlets, as opposed to working at the article level (we also modeled factuality of reporting, but as a separate task without trying multi-task learning). In addition to the text of the articles published by the target news medium, we used features extracted from its corresponding Wikipedia page and Twitter profile, as well as analysis of its URL structure and traffic information about it from Alexa rank. In the present work, we use a similar set of features, but we treat the problem as one of ordinal regression. Moreover, we model the political ideology and the factuality of reporting jointly in a multi-task learning setup, using several auxiliary tasks.

\paragraph{Multitask Ordinal Regression}

\emph{Ordinal regression} is well-studied and is commonly used for text classification on an ordinal scale, e.g.,~for  sentiment analysis on a 5-point scale \cite{he2016yzu,rosenthal2017semeval}.
However, \emph{multi-task ordinal regression} remains an understudied problem. 

\newcite{yu2006collaborative} proposed a Bayesian framework for collaborative ordinal regression, and demonstrated that modeling multiple ordinal regression tasks outperforms single-task models.

\noindent \newcite{walecki2016copula} were interested in jointly predicting facial action units and their intensity level.
They argued that, due to the high number of classes, modeling these tasks independently would be inefficient.
Thus, they proposed the \emph{copula ordinal regression} model for multi-task learning and demonstrated that it can outperform various single-task setups.
We use this model in our experiments below.

\newcite{balikas2017multitask} used multi-task ordinal regression for the task of fine-grained sentiment analysis.
In particular, they introduced an auxiliary coarse-grained task on a 3-point scale, and demonstrated that it can improve the results for sentiment analysis on the original 5-point scale.
Inspired by this, below we experiment with different granularity for political bias; however, we explore a larger space of possible auxiliary tasks.

\section{Method}\label{sec:method}

\paragraph{Copula Ordinal Regression}\label{subsec:cor}

We use the \emph{Copula Ordinal Regression} ({\sc COR}) model, which was originally proposed by~\newcite{walecki2016copula} to estimate the intensities of facial action units (AUs).
The model uses copula functions and conditional random fields (CRFs) to approximates the learning of the joint probability distribution function (PDF) of the facial AUs (random variables), using the bi-variate joint distributions capturing dependencies between AU pairs.
It was motivated by the fact that
\Ni many facial AUs co-exist with different levels of intensity, 
\Nii some AUs co-occur more often than others, and
\Niii some AUs depend on the intensity of other units.

\begin{table*}[tbh]
\centering
\scalebox{0.76}{
\begin{tabular}{ l|l|l|l|l|l}
\toprule
\bf Name & \bf URL & \bf Bias & \bf Factuality & \bf Twitter Handle & \bf Wikipedia page \\ \midrule
London Web News & \url{londonwebnews.com} & Extreme Left & Low & @londonwebnews & N/A \\
Daily Mirror & \url{www.mirror.co.uk} & Left & Mixed & @DailyMirror & \url{~/Daily_Mirror} \\
NBC News & \url{www.nbcnews.com} & Center-Left & High & @nbcnews & \url{~/NBC_News} \\
Associated Press & \url{apnews.com} & Center & Very High & @apnews & \url{~/Associated_Press} \\
Gulf News & \url{gulfnews.com} & Center-Right & High & @gulf\_news & \url{~/Gulf\_News} \\
Russia Insider & \url{russia-insider.com} & Right & Mixed & @russiainsider & \url{~/Russia_Insider} \\
Breitbart & \url{www.breitbart.com} & Extreme Right & Low & @BreitbartNews & \url{~/Breitbart_News} \\
\bottomrule
\end{tabular}}
\caption{Examples of media and their labels for bias and factuality of reporting derived from MBFC.\label{tab:examples}}
\end{table*}

We can draw an analogy between modeling facial AUs and modeling news media, where each medium expresses a particular bias (political ideology) and can also be associated with a particular level of factuality.
Therefore, bias and factuality can be analogous to the facial AUs in~\cite{walecki2016copula}, and represent two aspects of news reporting, each being modeled on a multi-point ordinal scale.
In particular, we model bias on a 7-point scale ({\it extreme-left}, {\it left}, {\it center-left}, {\it center}, {\it center-right}, {\it right}, and {\it extreme-right}), and
factuality on a 3-point scale ({\it low}, {\it mixed}, and {\it high}).

In our case, we train the {\sc COR} model to predict the joint PDF between political bias and factuality of reporting.
This could potentially work well given the inherent inter-dependency between the two tasks as we have seen on Figure~\ref{fig:mapping}.

\paragraph{Auxiliary Tasks}

We use a variety of auxiliary tasks, derived from the bias labels.
This includes converting the 7-point scale to \Ni 5-point and 3-point scales, similarly to~\cite{balikas2017multitask}, and to \Nii a 2-point scale in two ways to model extreme partisanship, and centrality.
Here is the list of the auxiliary tasks we use with precise definition of the label mappings:

\begin{itemize}[noitemsep,topsep=0pt]
  \item \textbf{Bias5-way:} Predict bias on a 5-pt scale; 1:\emph{extreme-left}, 2:\emph{left}, 3:\emph{\{center-left, center, center-right\}}, 4:{\it right}, and 5:{\it extreme-right}.
  \item \textbf{Bias3-way:} Predict bias on a 3-pt scale; 1:{\it \{extreme-left, left\}}, 2:{\it \{center-left, center, center-right\}}, and 3:{\it \{right, extreme-right\}}.
  \item \textbf{Bias-extreme:} Predict extreme vs. non-extreme partisanship on a 2-pt scale; 1:{\it \{extreme-left, extreme-right\}}, 2:{\it \{left, center-left, center, center-right, right\}}.
  \item \textbf{Bias-center:} Predict center vs. non-center political ideology on a 2-pt scale, ignoring polarity: 1:{\it \{extreme-left, left, right, extreme-right\}}, 2:{\it \{center-left, center, center-right\}}.
\end{itemize}

\paragraph{Features}

We used the features from~\citep{baly2018predicting}\footnote{\url{https://github.com/ramybaly/News-Media-Reliability}}.
We gathered a sample of articles from the target medium, and we calculated features such as POS tags, linguistic cues, sentiment scores, complexity, morality, as well as embeddings. We also used the Wikipedia page of the medium (if any) to generate document embedding. Then, we collected metadata from the medium's Twitter account (if any), e.g., whether is is verified, number of followers, whether the URL in the Twitter page matches the one of the medium. Finally, we added Web-based features that (\textit{i}) model the orthographic structure of the medium's URL address, and (\textit{ii}) analyze the Web-traffic information about the medium's website, as found in Alexa rank.\footnote{\url{https://www.alexa.com/siteinfo}}

\section{Experiments and Evaluation}\label{sec:experiments}

\paragraph{Data}\label{subsec:data}
We used the MBFC dataset~\cite{baly2018predicting} that has 1,066 news media manually annotated for factuality (3-pt scale: \emph{high}, \emph{mixed}, \emph{low}) and political bias (7-pt scale: from \emph{extreme-left} to \emph{extreme-right}).
This dataset was annotated by volunteers using a detailed methodology\footnote{For details, see \url{https://mediabiasfactcheck.com/methodology/}} that is designed to guarantee annotation objectivity.

\noindent Furthermore, readers can provide their own feedback on existing annotations, and in case of a large discrepancy, annotation is adjusted after a thorough review.
Therefore, we believe the annotation quality is good enough to experiment with.
We noticed that 117 media had \emph{low} factuality because they publish \emph{satire} and \emph{pseudo-science}, neither of which has a political perspective.
Since we are interested in modeling the relation between factuality and bias, we excluded those websites, thus ending up with 949 news media.
Some examples from this dataset are shown in Table~\ref{tab:examples} with both factuality and bias labels, in addition to their corresponding Twitter handles and Wikipedia pages. Overall, 64\% of the media in our dataset have Wikipedia pages, and 65\% have Twitter accounts.
Table~\ref{tab:stats} further provides detailed statistics about the label distribution in the MBFC dataset.

\begin{table}[tbh]
\centering
\small
\begin{tabular}{ll|lr}
\toprule
\multicolumn{2}{c|}{\bf Factuality} & \multicolumn{2}{c}{\bf Bias} \\ \midrule
Low   & 198 & Extreme-Left  & 23 \\
Mixed & 282 & Left          & 151 \\
High  & 469 & Center-Left   & 200 \\
      &     & Center        & 139 \\
      &     & Center-Right  & 105 \\
      &     & Right         & 164 \\
      &     & Extreme-Right & 167 \\ \bottomrule
\end{tabular}
\caption{Labels counts in the MBFC dataset that we used in our experiments.\label{tab:stats}}
\end{table}

\begin{table*}[tbh]
\centering
\small
\begin{tabular}{l|c:c|c:c}
\toprule
 & \multicolumn{2}{c|}{\bf Factuality} & \multicolumn{2}{c}{\bf Bias}\\ \midrule
{\bf Auxiliary Tasks} & {\bf \textcolor{white}{M}MAE\textcolor{white}{M}} & {\bf \textcolor{white}{M}MAE$^M$\textcolor{white}{M}} & {\bf MAE} & {\bf MAE$^M$}\\ \midrule
(None) majority class\dotfill & 0.714 & 1.000 & 1.798 & 1.857\\
(None) single-task {\sc COR}\dotfill & 0.514 & 0.567 & 1.582 & 1.728\\ \midrule
$+$bias\dotfill & 0.526 & 0.566 & -- & --\\
$+$factuality\dotfill & -- & -- & 1.584 & 1.695\\
$+$bias5-way\dotfill & 0.495 & 0.541 & 1.504\hspace{0.25cm}{\it(1.485)} & 1.627\hspace{0.25cm}{\it(1.647)}\\
$+$bias3-way\dotfill & 0.497 & 0.548 & 1.528\hspace{0.25cm}{\it(1.498)} & 1.658\hspace{0.25cm}{\it(1.654)}\\
$+$bias-center\dotfill & 0.509 & 0.561 & 1.594\hspace{0.25cm}{\it(1.535)} & 1.745\hspace{0.25cm}{\it(1.695)}\\
$+$bias-extreme\dotfill & 0.498 & 0.550 & 1.584\hspace{0.25cm}{\it(1.558)} & 1.743\hspace{0.25cm}{\it(1.726)}\\
$+$bias5-way$+$bias3-way\dotfill & 0.493 & 0.541 & 1.479\hspace{0.25cm}{\it({\bf1.475})} & 1.637\hspace{0.25cm}{\it ({\bf1.623})}\\
$+$bias-center$+$bias-extreme\dotfill & {\bf0.481} & {\bf0.529} & 1.563\hspace{0.25cm}{\it(1.526)} & 1.714\hspace{0.25cm}{\it (1.672)}\\
$+$bias5-way$+$bias3-way$+$bias-center$+$bias-extreme\dotfill & 0.485 & 0.537 & 1.513\hspace{0.25cm}{\it(1.504)} & 1.665\hspace{0.25cm}{\it(1.677)}\\
\bottomrule
\end{tabular}
\caption{Evaluating the copula ordinal regression model trained to jointly model the main task ({\it shown in the columns}) and different auxiliary tasks ({\it shown in the rows}). The results in parentheses correspond to the case when factuality is added as an additional auxiliary task (only applicable when the main task is bias prediction).}\label{tab:results}
\end{table*}

\paragraph{Experimental Setup}\label{subsec:setup}

We used the implementation\footnote{\url{https://github.com/RWalecki/copula_ordinal_regression}} of the Copula Ordinal Regression ({\sc COR}) model as described in \cite{walecki2016copula}.
In our experiments, we used 5-fold cross-validation, where for each fold we split the training dataset into a training part and a validation part, and we used the latter to fine-tune the model's hyper-parameters, optimizing for Mean Absolute Error (MAE). MAE is an appropriate evaluation measure given the ordinal nature of the tasks.

\noindent These hyper-parameters include the copula function (\emph{Gumbel} vs. \emph{Frank}), the marginal distribution (\emph{normal} vs. \emph{sigmoid}), the number of training iterations, the optimizer (\emph{gradient descent}, \emph{BFGS}), and the connection density of the CRFs.
We report both MAE and MAE$^M$, which is a variant of MAE that is more robust to class imbalance.
See \cite{Baccianella:2009qd,rosenthal-farra-nakov:2017:SemEval} for more details about MAE$^{M}$ vs. MAE.
We compare the results to two baselines: \Ni majority class, and \Nii single-task ordinal regression.

\paragraph{Results and Discussion}

Table~\ref{tab:results} shows the evaluation results for the {\sc COR} model when trained to jointly model the main task (\emph{shown in the columns}) using combinations of auxiliary tasks (\emph{shown in the rows}).
We can see that the single-task ordinal regression model performs much better than the majority class baseline based on both evaluation measures.
We can further see that the performance on the main task improves when jointly modeling several auxiliary tasks.
This improvement depends on the auxiliary tasks in use.

For factuality prediction, it turns out that the combination of {\it bias-center$+$bias-extreme} yields the best overall MAE of 0.481.
This makes sense and aligns well with the intuition that knowing whether a medium is centric or hyper-partisan is important to predict the factuality of its reporting.
For instance, a news medium without a political ideology tends to be more trustworthy compared to an extremely biased one, regardless of their polarity (left or right), as we should expect based on the data distribution shown in Figure~\ref{fig:mapping} above. 

For bias prediction (at a 7-point left-to-right scale), a joint model that uses political bias at different levels of granularity (5-point and 3-point) as auxiliary tasks yields the best overall MAE of 1.479.
This means that jointly modeling bias with the same information at coarser levels of granularity, i.e.,~adding 3-point and 5-point as auxiliary tasks, reduces the number of gross mistakes. 

\noindent E.g.,~predicting {\it extreme-left} instead of {\it extreme-right}, since the model is encouraged by the auxiliary tasks to learn the correct polarity, regardless of its intensity.
We can see that {\it factuality} is not very useful as an auxiliary task by itself (MAE=1.584 and MAE$^M$=1.695).
In other words, a medium with low factuality could be extremely biased to either the right or to the left.
Therefore, relying on {\it factuality} alone to predict bias might introduce severe errors, e.g., confusing extreme-left with extreme-right, thus leading to higher MAE scores.
This can be remedied by adding {\it factuality} to the mix of other auxiliary tasks to model the main task (7-point bias prediction).
The results of these experiments, shown in parentheses in Table~\ref{tab:results}, indicate that adding {\it factuality} to any combination of auxiliary tasks consistently yields lower MAE scores.
In particular, modeling the combination of {\it factuality$+$bias5-way$+$bias3-way} yields the best results (MAE=1.475 and MAE$^M$=1.623).
This result indicates that {\it factuality} provides complementary information that can help predict bias.

We ran a two-tailed t-test for statistical significance, which is suitable for an evaluation measure such as MAE, to confirm the improvements that were introduced by the multi-task setup.
We found that the best models (shown in bold in Table~\ref{tab:results}) outperformed both the corresponding majority class baselines with a p-value $\leq$ 0.001, and the corresponding single-task ordinal regression baselines with a p-value $\leq$ 0.02.

Finally, we compared the above results to our previous work \cite{baly2018predicting} by independently training a Support Vector Machine (SVM) classifier for each task, using the same features.

\noindent The resulting MAE was 0.450 for factuality and 1.184 for bias prediction, which is slightly better then our results (yet, very comparable for factuality).
However, our goal here is to emphasize the advantages of modeling the two tasks jointly.

\section{Conclusion and Future Work}\label{sec:conclusion}

We have presented a multi-task ordinal regression framework for jointly predicting
trustworthiness and political ideology of news media sources, using several auxiliary tasks, e.g.,~based on a coarser-grained scales or modeling extreme partisanship.
Overall, we have observed sizable performance gains in terms of reduced MAE by the multi-task ordinal regression models over single-task models for each of the two individual tasks.

In future work, we want to try more auxiliary tasks, and to experiment with other languages.
We further plan to go beyond \emph{left vs. right}, which is not universal and can exhibit regional specificity \cite{10.2307/27798525}, and to model other kinds of biases, e.g.,~\emph{eurosceptic vs. europhile}, \emph{nationalist} vs. \emph{globalist}, \emph{islamist vs. secular}, etc.

\section*{Acknowledgments}

This research is part of the Tanbih project,\footnote{\url{http://tanbih.qcri.org/}} which aims to limit the effect of ``fake news'', propaganda and media bias by making users aware of what they are reading. The project is developed in collaboration between the MIT Computer Science and Artificial Intelligence Laboratory (CSAIL) and the Qatar Computing Research Institute (QCRI), HBKU.

\bibliography{references}

\begin{thebibliography}{43}
\expandafter\ifx\csname natexlab\endcsname\relax\def\natexlab#1{#1}\fi

\bibitem[{Baccianella et~al.(2009)Baccianella, Esuli, and
  Sebastiani}]{Baccianella:2009qd}
Stefano Baccianella, Andrea Esuli, and Fabrizio Sebastiani. 2009.
\newblock Evaluation measures for ordinal regression.
\newblock In \emph{Proceedings of the 9th IEEE International Conference on
  Intelligent Systems Design and Applications}, ISDA~'09, pages 283--287, Pisa,
  Italy.

\bibitem[{Baeza-Yates(2018)}]{Baeza-Yates:2018:BW:3229066.3209581}
Ricardo Baeza-Yates. 2018.
\newblock \href {https://doi.org/10.1145/3209581} {Bias on the web}.
\newblock \emph{Commun. ACM}, 61(6):54--61.

\bibitem[{Balikas et~al.(2017)Balikas, Moura, and Amini}]{balikas2017multitask}
Georgios Balikas, Simon Moura, and Massih-Reza Amini. 2017.
\newblock Multitask learning for fine-grained {T}witter sentiment analysis.
\newblock In \emph{Proceedings of the 40th International ACM SIGIR Conference
  on Research and Development in Information Retrieval}, SIGIR~'17, pages
  1005--1008, Tokyo, Japan.

\bibitem[{Baly et~al.(2018)Baly, Karadzhov, Alexandrov, Glass, and
  Nakov}]{baly2018predicting}
Ramy Baly, Georgi Karadzhov, Dimitar Alexandrov, James Glass, and Preslav
  Nakov. 2018.
\newblock Predicting factuality of reporting and bias of news media sources.
\newblock In \emph{Proceedings of the Conference on Empirical Methods in
  Natural Language Processing}, EMNLP~'18, pages 3528--3539, Brussels, Belgium.

\bibitem[{Barr\'{o}n-Cede\~no et~al.(2019{\natexlab{a}})Barr\'{o}n-Cede\~no,
  Da~San~Martino, Jaradat, and Nakov}]{AAAI2019:proppy}
Alberto Barr\'{o}n-Cede\~no, Giovanni Da~San~Martino, Israa Jaradat, and
  Preslav Nakov. 2019{\natexlab{a}}.
\newblock Proppy: A system to unmask propaganda in online news.
\newblock In \emph{Proceedings of the Thirty-Third AAAI Conference on
  Artificial Intelligence}, AAAI'19, Honolulu, HI, USA.

\bibitem[{Barr\'{o}n-Cede\~no et~al.(2019{\natexlab{b}})Barr\'{o}n-Cede\~no,
  Da~San~Martino, Jaradat, and Nakov}]{Barron:19}
Alberto Barr\'{o}n-Cede\~no, Giovanni Da~San~Martino, Israa Jaradat, and
  Preslav Nakov. 2019{\natexlab{b}}.
\newblock Proppy: Organizing news coverage on the basis of their propagandistic
  content.
\newblock \emph{Information Processing and Management}.

\bibitem[{Brill(2001)}]{brill2001online}
Ann~M Brill. 2001.
\newblock Online journalists embrace new marketing function.
\newblock \emph{Newspaper Research Journal}, 22(2):28.

\bibitem[{Chen et~al.(2013)Chen, Wu, Srinivasan, and
  Zhang}]{Chen:2013:BIW:2492517.2492637}
Cheng Chen, Kui Wu, Venkatesh Srinivasan, and Xudong Zhang. 2013.
\newblock Battling the {I}nternet {W}ater {A}rmy: detection of hidden paid
  posters.
\newblock In \emph{Proceedings of the 2013 IEEE/ACM International Conference on
  Advances in Social Networks Analysis and Mining}, ASONAM '13, pages 116--120,
  Niagara, Canada.

\bibitem[{Darwish et~al.(2017)Darwish, Alexandrov, Nakov, and
  Mejova}]{SeminarUsers2017}
Kareem Darwish, Dimitar Alexandrov, Preslav Nakov, and Yelena Mejova. 2017.
\newblock Seminar users in the {A}rabic {T}witter sphere.
\newblock In \emph{Proceedings of the 9th International Conference on Social
  Informatics}, SocInfo~'17, pages 91--108, Oxford, UK.

\bibitem[{De~Sarkar et~al.(2018)De~Sarkar, Yang, and
  Mukherjee}]{desarkar-yang-mukherjee:2018:C18-1}
Sohan De~Sarkar, Fan Yang, and Arjun Mukherjee. 2018.
\newblock Attending sentences to detect satirical fake news.
\newblock In \emph{Proceedings of the 27th International Conference on
  Computational Linguistics}, COLING~'18, pages 3371--3380, Santa Fe, NM, USA.

\bibitem[{Dong et~al.(2015)Dong, Gabrilovich, Murphy, Dang, Horn, Lugaresi,
  Sun, and Zhang}]{Dong:2015:KTE:2777598.2777603}
Xin~Luna Dong, Evgeniy Gabrilovich, Kevin Murphy, Van Dang, Wilko Horn, Camillo
  Lugaresi, Shaohua Sun, and Wei Zhang. 2015.
\newblock Knowledge-based trust: Estimating the trustworthiness of web sources.
\newblock \emph{Proc. VLDB Endow.}, 8(9):938--949.

\bibitem[{Finberg et~al.(2002)Finberg, Stone, and Lynch}]{finberg2002digital}
Howard Finberg, Martha~L Stone, and Diane Lynch. 2002.
\newblock Digital journalism credibility study.
\newblock \emph{Online News Association. Retrieved November}, 3:2003.

\bibitem[{Gerrish and Blei(2011)}]{Gerrish:2011:PLR:3104482.3104544}
Sean~M. Gerrish and David~M. Blei. 2011.
\newblock Predicting legislative roll calls from text.
\newblock In \emph{Proceedings of the 28th International Conference on
  International Conference on Machine Learning}, ICML~'11, pages 489--496,
  Bellevue, Washington, USA.

\bibitem[{Hardalov et~al.(2016)Hardalov, Koychev, and Nakov}]{Hardalov2016}
Momchil Hardalov, Ivan Koychev, and Preslav Nakov. 2016.
\newblock In search of credible news.
\newblock In \emph{Proceedings of the 17th International Conference on
  Artificial Intelligence: Methodology, Systems, and Applications}, AIMSA~'16,
  pages 172--180, Varna, Bulgaria.

\bibitem[{He et~al.(2016)He, Yu, Yang, Lai, and Liu}]{he2016yzu}
Yunchao He, Liang-Chih Yu, Chin-Sheng Yang, K~Robert Lai, and Weiyi Liu. 2016.
\newblock {YZU-NLP} team at semeval-2016 task 4: Ordinal sentiment
  classification using a recurrent convolutional network.
\newblock In \emph{Proceedings of the 10th International Workshop on Semantic
  Evaluation}, SemEval~'16, pages 251--255, San Diego, CA, USA.

\bibitem[{Horne et~al.(2018{\natexlab{a}})Horne, Khedr, and
  Adali}]{DBLP:journals/corr/abs-1803-10124}
Benjamin Horne, Sara Khedr, and Sibel Adali. 2018{\natexlab{a}}.
\newblock Sampling the news producers: A large news and feature data set for
  the study of the complex media landscape.
\newblock In \emph{Proceedings of the Twelfth International Conference on Web
  and Social Media}, ICWSM~'18, pages 518--527, Stanford, CA, USA.

\bibitem[{Horne et~al.(2018{\natexlab{b}})Horne, Dron, Khedr, and
  Adali}]{Horne:2018:ANL:3184558.3186987}
Benjamin~D. Horne, William Dron, Sara Khedr, and Sibel Adali.
  2018{\natexlab{b}}.
\newblock Assessing the news landscape: A multi-module toolkit for evaluating
  the credibility of news.
\newblock In \emph{Proceedings of the The Web Conference}, WWW~'18, pages
  235--238, Lyon, France.

\bibitem[{Iyyer et~al.(2014)Iyyer, Enns, Boyd-Graber, and Resnik}]{P14-1105}
Mohit Iyyer, Peter Enns, Jordan Boyd-Graber, and Philip Resnik. 2014.
\newblock Political ideology detection using recursive neural networks.
\newblock In \emph{Proceedings of the 52nd Annual Meeting of the Association
  for Computational Linguistics}, pages 1113--1122, Baltimore, MD, USA.

\bibitem[{Karadzhov et~al.(2017)Karadzhov, Gencheva, Nakov, and
  Koychev}]{RANLP2017:clickbait}
Georgi Karadzhov, Pepa Gencheva, Preslav Nakov, and Ivan Koychev. 2017.
\newblock We built a fake news \& click-bait filter: What happened next will
  blow your mind!
\newblock In \emph{Proceedings of the International Conference on Recent
  Advances in Natural Language Processing}, RANLP~'17, pages 334--343, Varna,
  Bulgaria.

\bibitem[{Kulkarni et~al.(2018)Kulkarni, Ye, Skiena, and
  Wang}]{kulkarni2018multi}
Vivek Kulkarni, Junting Ye, Steven Skiena, and William~Yang Wang. 2018.
\newblock Multi-view models for political ideology detection of news articles.
\newblock In \emph{Proceedings of the Conference on Empirical Methods in
  Natural Language Processing}, EMNLP~'18, pages 3518--3527, Brussels, Belgium.

\bibitem[{Lazer et~al.(2018)Lazer, Baum, Benkler, Berinsky, Greenhill, Menczer,
  Metzger, Nyhan, Pennycook, Rothschild, Schudson, Sloman, Sunstein, Thorson,
  Watts, and Zittrain}]{Lazer1094}
David~M.J. Lazer, Matthew~A. Baum, Yochai Benkler, Adam~J. Berinsky, Kelly~M.
  Greenhill, Filippo Menczer, Miriam~J. Metzger, Brendan Nyhan, Gordon
  Pennycook, David Rothschild, Michael Schudson, Steven~A. Sloman, Cass~R.
  Sunstein, Emily~A. Thorson, Duncan~J. Watts, and Jonathan~L. Zittrain. 2018.
\newblock The science of fake news.
\newblock \emph{Science}, 359(6380):1094--1096.

\bibitem[{Maity et~al.(2017)Maity, Chakraborty, Goyal, and
  Mukherjee}]{Maity:2017:DSS:3022198.3026360}
Suman~Kalyan Maity, Aishik Chakraborty, Pawan Goyal, and Animesh Mukherjee.
  2017.
\newblock {D}etection of sockpuppets in social media.
\newblock In \emph{Proceedings of the ACM Conference on Computer Supported
  Cooperative Work and Social Computing}, CSCW~'17, pages 243--246, Portland,
  OR, USA.

\bibitem[{Mihaylov et~al.(2018)Mihaylov, Mihaylova, Nakov, M\`{a}rquez,
  Georgiev, and Koychev}]{InternetResearchJournal:2018}
Todor Mihaylov, Tsvetomila Mihaylova, Preslav Nakov, Llu\'{i}s M\`{a}rquez,
  Georgi Georgiev, and Ivan Koychev. 2018.
\newblock The dark side of news community forums: Opinion manipulation trolls.
\newblock \emph{Internet Research}, 28(5):1292--1312.

\bibitem[{Mihaylov and Nakov(2016)}]{mihaylov-nakov:2016:P16-2}
Todor Mihaylov and Preslav Nakov. 2016.
\newblock {H}unting for troll comments in news community forums.
\newblock In \emph{Proceedings of the 54th Annual Meeting of the Association
  for Computational Linguistics}, ACL~'16, pages 399--405, Berlin, Germany.

\bibitem[{Mihaylova et~al.(2018)Mihaylova, Nakov, M\`{a}rquez,
  Barr\'on-Cede{\~n}o, Mohtarami, Karadjov, and Glass}]{AAAI2018:factchecking}
Tsvetomila Mihaylova, Preslav Nakov, Llu\'{i}s M\`{a}rquez, Alberto
  Barr\'on-Cede{\~n}o, Mitra Mohtarami, Georgi Karadjov, and James Glass. 2018.
\newblock Fact checking in community forums.
\newblock In \emph{Proceedings of the Thirty-Second AAAI Conference on
  Artificial Intelligence}, AAAI~'18, pages 879--886, New Orleans, LA, USA.

\bibitem[{Mukherjee and Weikum(2015)}]{mukherjee2015leveraging}
Subhabrata Mukherjee and Gerhard Weikum. 2015.
\newblock Leveraging joint interactions for credibility analysis in news
  communities.
\newblock In \emph{Proceedings of the 24th ACM International on Conference on
  Information and Knowledge Management}, CIKM~'15, pages 353--362, Melbourne,
  Australia.

\bibitem[{Nguyen et~al.(2018)Nguyen, Kharosekar, Lease, and
  Wallace}]{DBLP:conf/aaai/NguyenKLW18}
An~T. Nguyen, Aditya Kharosekar, Matthew Lease, and Byron~C. Wallace. 2018.
\newblock An interpretable joint graphical model for fact-checking from crowds.
\newblock In \emph{Proceedings of the Thirty-Second {AAAI} Conference on
  Artificial Intelligence}, AAAI~'18, New Orleans, LA, USA.

\bibitem[{Pan et~al.(2018)Pan, Pavlova, Li, Li, Li, and Liu}]{Pan:KG:2018}
Jeff~Z. Pan, Siyana Pavlova, Chenxi Li, Ningxi Li, Yangmei Li, and Jinshuo Liu.
  2018.
\newblock Content based fake news detection using knowledge graphs.
\newblock In \emph{Proceedings of the International Semantic Web Conference},
  ISWC~'18, Monterey, CA, USA.

\bibitem[{P\'{e}rez-Rosas et~al.(2018)P\'{e}rez-Rosas, Kleinberg, Lefevre, and
  Mihalcea}]{prezrosas-EtAl:2018:C18-1}
Ver\'{o}nica P\'{e}rez-Rosas, Bennett Kleinberg, Alexandra Lefevre, and Rada
  Mihalcea. 2018.
\newblock Automatic detection of fake news.
\newblock In \emph{Proceedings of the 27th International Conference on
  Computational Linguistics}, COLING~'18, pages 3391--3401, Santa Fe, NM, USA.

\bibitem[{Pitoura et~al.(2018)Pitoura, Tsaparas, Flouris, Fundulaki, Papadakos,
  Abiteboul, and Weikum}]{Pitoura:2018:MBO:3186549.3186553}
Evaggelia Pitoura, Panayiotis Tsaparas, Giorgos Flouris, Irini Fundulaki,
  Panagiotis Papadakos, Serge Abiteboul, and Gerhard Weikum. 2018.
\newblock On measuring bias in online information.
\newblock \emph{SIGMOD Rec.}, 46(4):16--21.

\bibitem[{Popat et~al.(2016)Popat, Mukherjee, Str\"{o}tgen, and
  Weikum}]{popat2016credibility}
Kashyap Popat, Subhabrata Mukherjee, Jannik Str\"{o}tgen, and Gerhard Weikum.
  2016.
\newblock Credibility assessment of textual claims on the web.
\newblock In \emph{Proceedings of the 25th ACM International on Conference on
  Information and Knowledge Management}, CIKM~'16, pages 2173--2178,
  Indianapolis, IN, USA.

\bibitem[{Popat et~al.(2017)Popat, Mukherjee, Str\"{o}tgen, and
  Weikum}]{Popat:2017:TLE:3041021.3055133}
Kashyap Popat, Subhabrata Mukherjee, Jannik Str\"{o}tgen, and Gerhard Weikum.
  2017.
\newblock Where the truth lies: Explaining the credibility of emerging claims
  on the {W}eb and social media.
\newblock In \emph{Proceedings of the 26th International Conference on World
  Wide Web Companion}, WWW~'17, pages 1003--1012, Perth, Australia.

\bibitem[{Popat et~al.(2018)Popat, Mukherjee, Str\"{o}tgen, and
  Weikum}]{Popat:2018:CCL:3184558.3186967}
Kashyap Popat, Subhabrata Mukherjee, Jannik Str\"{o}tgen, and Gerhard Weikum.
  2018.
\newblock {CredEye}: A credibility lens for analyzing and explaining
  misinformation.
\newblock In \emph{Proceedings of The Web Conference 2018}, WWW~'18, pages
  155--158, Lyon, France.

\bibitem[{Potthast et~al.(2018)Potthast, Kiesel, Reinartz, Bevendorff, and
  Stein}]{DBLP:journals/corr/PotthastKRBS17}
Martin Potthast, Johannes Kiesel, Kevin Reinartz, Janek Bevendorff, and Benno
  Stein. 2018.
\newblock A stylometric inquiry into hyperpartisan and fake news.
\newblock In \emph{Proceedings of the 56th Annual Meeting of the Association
  for Computational Linguistics}, ACL~'18, pages 231--240, Melbourne,
  Australia.

\bibitem[{Preo{\c{t}}iuc-Pietro et~al.(2017)Preo{\c{t}}iuc-Pietro, Liu,
  Hopkins, and Ungar}]{P17-1068}
Daniel Preo{\c{t}}iuc-Pietro, Ye~Liu, Daniel Hopkins, and Lyle Ungar. 2017.
\newblock Beyond binary labels: Political ideology prediction of {T}witter
  users.
\newblock In \emph{Proceedings of the 55th Annual Meeting of the Association
  for Computational Linguistics (Volume 1: Long Papers)}, ACL~'17, pages
  729--740, Vancouver, Canada.

\bibitem[{Rashkin et~al.(2017)Rashkin, Choi, Jang, Volkova, and
  Choi}]{rashkin-EtAl:2017:EMNLP2017}
Hannah Rashkin, Eunsol Choi, Jin~Yea Jang, Svitlana Volkova, and Yejin Choi.
  2017.
\newblock Truth of varying shades: Analyzing language in fake news and
  political fact-checking.
\newblock In \emph{Proceedings of the 2017 Conference on Empirical Methods in
  Natural Language Processing}, EMNLP~'17, pages 2931--2937, Copenhagen,
  Denmark.

\bibitem[{Rosenthal et~al.(2017{\natexlab{a}})Rosenthal, Farra, and
  Nakov}]{rosenthal2017semeval}
Sara Rosenthal, Noura Farra, and Preslav Nakov. 2017{\natexlab{a}}.
\newblock {SemEval}-2017 task 4: Sentiment analysis in {T}witter.
\newblock In \emph{Proceedings of the 11th International Workshop on Semantic
  Evaluation}, SemEval~'17, pages 502--518, Vancouver, Canada.

\bibitem[{Rosenthal et~al.(2017{\natexlab{b}})Rosenthal, Farra, and
  Nakov}]{rosenthal-farra-nakov:2017:SemEval}
Sara Rosenthal, Noura Farra, and Preslav Nakov. 2017{\natexlab{b}}.
\newblock {SemEval}-2017 task 4: Sentiment analysis in {T}witter.
\newblock In \emph{Proceedings of the 11th International Workshop on Semantic
  Evaluation}, SemEval~'17, pages 502--518, Vancouver, Canada.

\bibitem[{Sim et~al.(2013)Sim, Acree, Gross, and Smith}]{D13-1010}
Yanchuan Sim, Brice D.~L. Acree, Justin~H. Gross, and Noah~A. Smith. 2013.
\newblock Measuring ideological proportions in political speeches.
\newblock In \emph{Proceedings of the 2013 Conference on Empirical Methods in
  Natural Language Processing}, EMNLP~'13, pages 91--101, Seattle, WA, USA.

\bibitem[{Tavits and Letki(2009)}]{10.2307/27798525}
Margit Tavits and Natalia Letki. 2009.
\newblock When left is right: Party ideology and policy in {Post-Communist
  Europe}.
\newblock \emph{The American Political Science Review}, 103(4):555--569.

\bibitem[{Vosoughi et~al.(2018)Vosoughi, Roy, and Aral}]{Vosoughi1146}
Soroush Vosoughi, Deb Roy, and Sinan Aral. 2018.
\newblock The spread of true and false news online.
\newblock \emph{Science}, 359(6380):1146--1151.

\bibitem[{Walecki et~al.(2016)Walecki, Rudovic, Pavlovic, and
  Pantic}]{walecki2016copula}
Robert Walecki, Ognjen Rudovic, Vladimir Pavlovic, and Maja Pantic. 2016.
\newblock Copula ordinal regression for joint estimation of facial action unit
  intensity.
\newblock In \emph{Proceedings of the IEEE Conference on Computer Vision and
  Pattern Recognition}, pages 4902--4910.

\bibitem[{Yu et~al.(2006)Yu, Yu, Tresp, and Kriegel}]{yu2006collaborative}
Shipeng Yu, Kai Yu, Volker Tresp, and Hans-Peter Kriegel. 2006.
\newblock Collaborative ordinal regression.
\newblock In \emph{Proceedings of the 23rd international conference on Machine
  learning}, pages 1089--1096. ACM.

\end{thebibliography}


@InProceedings{peters2018deep,
  title       = {Deep Contextualized Word Representations},
  author      = {Peters, Matthew and Neumann, Mark and Iyyer, Mohit and Gardner, Matt and Clark, Christopher and Lee, Kenton and Zettlemoyer, Luke},
  booktitle   = {Proceedings of the 2018 Conference of the North American Chapter of the Association for Computational Linguistics: Human Language Technologies},
  series      = {NAACL-HLT~'18},
  NOmonth     = {June},
  address     = {New Orleans, LA, USA},
  NOpublisher = {Association for Computational Linguistics},
  pages       = {2227--2237},
  year        = {2018}
}

@InProceedings{mikolov2013distributed,
  title     = {Distributed representations of words and phrases and their compositionality},
  author    = {Mikolov, Tomas and Sutskever, Ilya and Chen, Kai and Corrado, Greg S and Dean, Jeff},
  booktitle = {Advances in neural information processing systems},
  pages     = {3111--3119},
  year      = {2013}
}

@InProceedings{walecki2016copula,
  title     = {Copula ordinal regression for joint estimation of facial action unit intensity},
  author    = {Walecki, Robert and Rudovic, Ognjen and Pavlovic, Vladimir and Pantic, Maja},
  booktitle = {Proceedings of the IEEE Conference on Computer Vision and Pattern Recognition},
  pages     = {4902--4910},
  year      = {2016}
}

@InProceedings{rashkin-EtAl:2017:EMNLP2017,
  author      = {Rashkin, Hannah  and  Choi, Eunsol  and  Jang, Jin Yea  and  Volkova, Svitlana  and  Choi, Yejin},
  title       = {Truth of Varying Shades: Analyzing Language in Fake News and Political Fact-Checking},
  booktitle   = {Proceedings of the 2017 Conference on Empirical Methods in Natural Language Processing},
  series      = {EMNLP~'17},
  NOmonth     = {September},
  year        = {2017},
  address     = {Copenhagen, Denmark},
  NOpublisher = {Association for Computational Linguistics},
  pages       = {2931--2937},
  NOurl       = {https://www.aclweb.org/anthology/D17-1317}
}

@article{Shu:2017:FND:3137597.3137600,
 author     = {Shu, Kai and Sliva, Amy and Wang, Suhang and Tang, Jiliang and Liu, Huan},
 title      = {Fake News Detection on Social Media: A Data Mining Perspective},
 journal    = {SIGKDD Explor. Newsl.},
 issue_date = {June 2017},
 volume     = {19},
 number     = {1},
 month      = sep,
 year       = {2017},
 issn       = {1931-0145},
 pages      = {22--36},
 numpages   = {15},
 url        = {http://doi.acm.org/10.1145/3137597.3137600},
 doi        = {10.1145/3137597.3137600},
 acmid      = {3137600},
 publisher  = {ACM},
 address    = {New York, NY, USA},
} 

@InProceedings{DBLP:journals/corr/PotthastKRBS17,
  author      = {Potthast, Martin  and  Kiesel, Johannes  and  Reinartz, Kevin  and  Bevendorff, Janek  and  Stein, Benno},
  title       = {A Stylometric Inquiry into Hyperpartisan and Fake News},
  booktitle   = {Proceedings of the 56th Annual Meeting of the Association for Computational Linguistics},
  series      = {ACL~'18},
  NOmonth     = {July},
  year        = {2018},
  address     = {Melbourne, Australia},
  NOpublisher = {Association for Computational Linguistics},
  pages       = {231--240},
  NOurl       = {http://www.aclweb.org/anthology/P18-1022}
}

@inproceedings{DBLP:journals/corr/abs-1803-10124,
  author    = {Horne, Benjamin and Khedr, Sara and Adali, Sibel},
  title     = {Sampling the News Producers: A Large News and Feature Data Set for the Study of the Complex Media Landscape},
  booktitle = {Proceedings of the Twelfth International Conference on Web and Social Media},
  series    = {ICWSM~'18},
  address   = {Stanford, CA, USA},
  pages     = {518--527},
  year      = {2018},
}

@article{DBLP:journals/corr/HorneA17,
  author        = {Horne, Benjamin and Adali, Sibel},
  title         = {This Just In: Fake News Packs a Lot in Title, Uses Simpler, Repetitive Content in Text Body, More Similar to Satire than Real News},
  journal       = {CoRR},
  volume        = {abs/1703.09398},
  year          = {2017},
  url           = {http://arxiv.org/abs/1703.09398},
  archivePrefix = {arXiv},
  eprint        = {1703.09398},
  timestamp     = {Wed, 07 Jun 2017 14:41:09 +0200},
  biburl        = {https://dblp.org/rec/bib/journals/corr/HorneA17},
  bibsource     = {dblp computer science bibliography, https://dblp.org}
}

@article{Vosoughi1146,
	author = {Vosoughi, Soroush and Roy, Deb and Aral, Sinan},
	title = {The spread of true and false news online},
	volume = {359},
	number = {6380},
	pages = {1146--1151},
	year = {2018},
	NOdoi = {10.1126/science.aap9559},
	NOpublisher = {American Association for the Advancement of Science},
	issn = {0036-8075},
	NOURL = {http://science.sciencemag.org/content/359/6380/1146},
	NOeprint = {http://science.sciencemag.org/content/359/6380/1146.full.pdf},
	journal = {Science}
}

@article{Lazer1094,
	author = {Lazer, David M.J. and Baum, Matthew A. and Benkler, Yochai and Berinsky, Adam J. and Greenhill, Kelly M. and Menczer, Filippo and Metzger, Miriam J. and Nyhan, Brendan and Pennycook, Gordon and Rothschild, David and Schudson, Michael and Sloman, Steven A. and Sunstein, Cass R. and Thorson, Emily A. and Watts, Duncan J. and Zittrain, Jonathan L.},
	title = {The science of fake news},
	volume = {359},
	number = {6380},
	pages = {1094--1096},
	year = {2018},
	NOdoi = {10.1126/science.aao2998},
	NOpublisher = {American Association for the Advancement of Science},
	issn = {0036-8075},
	NOURL = {http://science.sciencemag.org/content/359/6380/1094},
	NOeprint = {http://science.sciencemag.org/content/359/6380/1094.full.pdf},
	journal = {Science}
}

@inproceedings{ma2009identifying,
 author = {Ma, Justin and Saul, Lawrence K. and Savage, Stefan and Voelker, Geoffrey M.},
 title = {Identifying Suspicious {URLs}: An Application of Large-scale Online Learning},
 booktitle = {Proceedings of the 26th Annual International Conference on Machine Learning},
 series = {ICML~'09},
 year = {2009},
 isbn = {978-1-60558-516-1},
 address = {Montreal, Canada},
 pages = {681--688},
 numpages = {8},
 NOurl = {http://doi.acm.org/10.1145/1553374.1553462},
 NOdoi = {10.1145/1553374.1553462},
 acmid = {1553462},
 NOpublisher = {ACM},
 NOaddress = {New York, NY, USA},
} 

@inproceedings{rosenthal2017semeval,
  author    = {Rosenthal, Sara  and  Farra, Noura  and  Nakov, Preslav},
  title     = {{SemEval}-2017 Task 4: Sentiment Analysis in {T}witter},
  booktitle = {Proceedings of the 11th International Workshop on Semantic Evaluation},
  series = {SemEval~'17},
  NOmonth     = {August},
  year      = {2017},
  address   = {Vancouver, Canada},
  NOpublisher = {Association for Computational Linguistics},
  pages     = {502--518},
  NOurl       = {http://www.aclweb.org/anthology/S17-2088}
}

@inproceedings{Baccianella:2009qd,
	Address = {Pisa, Italy},
	Author = {Baccianella, Stefano and Esuli, Andrea and Sebastiani, Fabrizio},
	Booktitle = {Proceedings of the 9th IEEE International Conference on Intelligent Systems Design and Applications},
    series = {ISDA~'09},
	Date-Added = {2015-07-01 12:48:04 +0000},
	Date-Modified = {2015-07-01 12:48:04 +0000},
	Pages = {283--287},
	Title = {Evaluation Measures for Ordinal Regression},
	Year = {2009}}

@inproceedings{Horne:2018:ANL:3184558.3186987,
 author = {Horne, Benjamin D. and Dron, William and Khedr, Sara and Adali, Sibel},
 title = {Assessing the News Landscape: A Multi-Module Toolkit for Evaluating the Credibility of News},
 booktitle = {Proceedings of the The Web Conference},
 series = {WWW~'18},
 year = {2018},
 isbn = {978-1-4503-5640-4},
 address = {Lyon, France},
 pages = {235--238},
 numpages = {4},
 NOurl = {https://doi.org/10.1145/3184558.3186987},
 NOdoi = {10.1145/3184558.3186987},
 acmid = {3186987},
 NOpublisher = {International World Wide Web Conferences Steering Committee},
 NOaddress = {Republic and Canton of Geneva, Switzerland},
 keywords = {content analysis, data exploration, fact-checking assistance, news credibility, toolkit},
}

@inproceedings{DBLP:conf/aaai/NguyenKLW18,
  author    = {An T. Nguyen and
               Aditya Kharosekar and
               Matthew Lease and
               Byron C. Wallace},
  NOeditor    = {Sheila A. McIlraith and
               Kilian Q. Weinberger},
  title     = {An Interpretable Joint Graphical Model for Fact-Checking From Crowds},
  booktitle = {Proceedings of the Thirty-Second {AAAI} Conference on Artificial Intelligence},
  series = {AAAI~'18},
  address = {New Orleans, LA, USA},
  NOpublisher = {{AAAI} Press},
  year      = {2018},
  NOurl       = {https://www.aaai.org/ocs/index.php/AAAI/AAAI18/paper/view/16673},
}

@inproceedings{Popat:2017:TLE:3041021.3055133,
 author = {Popat, Kashyap and Mukherjee, Subhabrata and Str\"{o}tgen, Jannik and Weikum, Gerhard},
 title = {Where the Truth Lies: Explaining the Credibility of Emerging Claims on the {W}eb and Social Media},
 booktitle = {Proceedings of the 26th International Conference on World Wide Web Companion},
 series = {WWW~'17},
 year = {2017},
 isbn = {978-1-4503-4914-7},
 address = {Perth, Australia},
 pages = {1003--1012},
 numpages = {10},
 NOurl = {https://doi.org/10.1145/3041021.3055133},
 NOdoi = {10.1145/3041021.3055133},
 acmid = {3055133},
 NOpublisher = {International World Wide Web Conferences Steering Committee},
 NOaddress = {Republic and Canton of Geneva, Switzerland},
 NOkeywords = {credibility analysis, rumor and hoax detection, text mining},
} 

@article{Dong:2015:KTE:2777598.2777603,
 author = {Dong, Xin Luna and Gabrilovich, Evgeniy and Murphy, Kevin and Dang, Van and Horn, Wilko and Lugaresi, Camillo and Sun, Shaohua and Zhang, Wei},
 title = {Knowledge-based Trust: Estimating the Trustworthiness of Web Sources},
 journal = {Proc. VLDB Endow.},
 issue_date = {May 2015},
 volume = {8},
 number = {9},
 month = may,
 year = {2015},
 issn = {2150-8097},
 pages = {938--949},
 numpages = {12},
 NOurl = {https://doi.org/10.14778/2777598.2777603},
 NOdoi = {10.14778/2777598.2777603},
 NOacmid = {2777603},
 NOpublisher = {VLDB Endowment},
} 

@inproceedings{DBLP:conf/www/ZhangRMASGAVLRB18,
  author    = {Amy X. Zhang and
               Aditya Ranganathan and
               Sarah Emlen Metz and
               Scott Appling and
               Connie Moon Sehat and
               Norman Gilmore and
               Nick B. Adams and
               Emmanuel Vincent and
               Jennifer Lee and
               Martin Robbins and
               Ed Bice and
               Sandro Hawke and
               David R. Karger and
               An Xiao Mina},
  NOeditor    = {Pierre{-}Antoine Champin and
               Fabien L. Gandon and
               Mounia Lalmas and
               Panagiotis G. Ipeirotis},
  title     = {A Structured Response to Misinformation: Defining and Annotating Credibility Indicators in News Articles},
  booktitle = {Proceedings of the The Web Conference (Companion)},
  series    = {WWW~'18 companion},
  address   = {Lyon, France},
  pages     = {603--612},
  NOpublisher = {{ACM}},
  year      = {2018},
  NOurl     = {http://doi.acm.org/10.1145/3184558.3188731},
  NOdoi     = {10.1145/3184558.3188731},
}

@article{DBLP:journals/tmm/JinCZZT17,
  author    = {Zhiwei Jin and
               Juan Cao and
               Yongdong Zhang and
               Jianshe Zhou and
               Qi Tian},
  title     = {Novel Visual and Statistical Image Features for Microblogs News Verification},
  journal   = {{IEEE} Trans. Multimedia},
  volume    = {19},
  number    = {3},
  pages     = {598--608},
  year      = {2017},
  url       = {https://doi.org/10.1109/TMM.2016.2617078},
  doi       = {10.1109/TMM.2016.2617078},
  timestamp = {Sun, 28 May 2017 13:24:19 +0200},
  biburl    = {https://dblp.org/rec/bib/journals/tmm/JinCZZT17},
  bibsource = {dblp computer science bibliography, https://dblp.org}
}

@inproceedings{Gupta:2013:FSC:2487788.2488033,
 author = {Gupta, Aditi and Lamba, Hemank and Kumaraguru, Ponnurangam and Joshi, Anupam},
 title = {Faking {S}andy: Characterizing and Identifying Fake Images on {T}witter During Hurricane {S}andy},
 booktitle = {Proceedings of the 22Nd International Conference on World Wide Web},
 series = {WWW~'13 Companion},
 year = {2013},
 isbn = {978-1-4503-2038-2},
 address = {Rio de Janeiro, Brazil},
 pages = {729--736},
 numpages = {8},
 NOurl = {http://doi.acm.org/10.1145/2487788.2488033},
 NOdoi = {10.1145/2487788.2488033},
 acmid = {2488033},
 NOpublisher = {ACM},
 NOaddress = {New York, NY, USA},
 NOkeywords = {online social networks},
} 

@inproceedings{Castillo:2011:ICT:1963405.1963500,
 author = {Castillo, Carlos and Mendoza, Marcelo and Poblete, Barbara},
 title = {Information Credibility on {T}witter},
 booktitle = {Proceedings of the 20th International Conference on World Wide Web},
 series = {WWW~'11},
 year = {2011},
 isbn = {978-1-4503-0632-4},
 address = {Hyderabad, India},
 pages = {675--684},
 numpages = {10},
 NOurl = {http://doi.acm.org/10.1145/1963405.1963500},
 NOdoi = {10.1145/1963405.1963500},
 acmid = {1963500},
 NOpublisher = {ACM},
 NOaddress = {New York, NY, USA},
 NOkeywords = {social media analytics, social media credibility, twitter},
} 

@InProceedings{P17-1066,
  author = 	"Ma, Jing
		and Gao, Wei
		and Wong, Kam-Fai",
  title = 	"Detect Rumors in Microblog Posts Using Propagation Structure via Kernel      Learning    ",
  booktitle = 	"Proceedings of the 55th Annual Meeting of the Association for      Computational Linguistics",
  series = {ACL~'17},
  year = 	"2017",
  NOpublisher = 	"Association for Computational Linguistics",
  pages = 	"708--717",
  address = 	"Vancouver, Canada",
  NOdoi = 	"10.18653/v1/P17-1066",
  NOurl = 	"http://www.aclweb.org/anthology/P17-1066"
}

@inproceedings{Jin:2016:NVE:3016100.3016318,
 author = {Jin, Zhiwei and Cao, Juan and Zhang, Yongdong and Luo, Jiebo},
 title = {News Verification by Exploiting Conflicting Social Viewpoints in Microblogs},
 booktitle = {Proceedings of the Thirtieth AAAI Conference on Artificial Intelligence},
 series = {AAAI~'16},
 year = {2016},
 address = {Phoenix, AZ, USA},
 pages = {2972--2978},
 numpages = {7},
 NOurl = {http://dl.acm.org/citation.cfm?id=3016100.3016318},
 acmid = {3016318},
 NOpublisher = {AAAI Press},
} 

@article{Li:2016:STD:2897350.2897352,
 author = {Li, Yaliang and Gao, Jing and Meng, Chuishi and Li, Qi and Su, Lu and Zhao, Bo and Fan, Wei and Han, Jiawei},
 title = {A Survey on Truth Discovery},
 journal = {SIGKDD Explor. Newsl.},
 issue_date = {December 2015},
 volume = {17},
 number = {2},
 month = feb,
 year = {2016},
 issn = {1931-0145},
 pages = {1--16},
 numpages = {16},
 NOurl = {http://doi.acm.org/10.1145/2897350.2897352},
 NOdoi = {10.1145/2897350.2897352},
 acmid = {2897352},
 NOpublisher = {ACM},
 NOaddress = {New York, NY, USA},
} 


@InProceedings{clef2018checkthat,
    author    = {Nakov, Preslav  and  Barr\'{o}n-Cede\~no, Alberto and Elsayed, Tamer and Suwaileh, Reem and M\`{a}rquez, Llu\'{i}s and Zaghouani, Wajdi and Gencheva, Pepa and Kyuchukov, Spas and Da San Martino, Giovanni},
    title     = {{CLEF}-2018 Lab on Automatic Identification and Verification of Claims in Political Debates},
    booktitle = {Working Notes of {CLEF} 2018 -- Conference and Labs of the Evaluation
               Forum},
    series    = {CLEF~'18},
    address   = {Avignon, France},
    NOmonth     = {September},
    year      = {2018}, 
}

@inproceedings{Agichtein:2008:FHC:1341531.1341557,
 author = {Agichtein, Eugene and Castillo, Carlos and Donato, Debora and Gionis, Aristides and Mishne, Gilad},
 title = {Finding High-quality Content in Social Media},
 booktitle = {Proceedings of the International Conference on Web Search and Data Mining},
 series = {WSDM '08},
 year = {2008},
 isbn = {978-1-59593-927-2},
 address = {Palo Alto, CA, USA},
 pages = {183--194},
 numpages = {12},
 acmid = {1341557},
 NOpublisher = {ACM},
 NOaddress = {New York, NY, USA},
 keywords = {community question answering, media, user interactions},
}

@article{Bu:2013:SPD:2400768.2401510,
 author = {Bu, Zhan and Xia, Zhengyou and Wang, Jiandong},
 title = {A sock puppet detection algorithm on virtual spaces},
 journal = {Know.-Based Syst.},
 issue_date = {2013},
 volume = {37},
 month = jan,
 year = {2013},
 issn = {0950-7051},
 pages = {366--377},
 numpages = {12},
 NOdoi = {10.1016/j.knosys.2012.08.016},
 acmid = {2401510},
 publisher = {Elsevier Science Publishers B. V.},
 address = {Amsterdam, The Netherlands, The Netherlands},
 keywords = {Authorship identification, Community detection, Hypothesis test, Sock puppet, Virtual spaces},
} 

@inproceedings{Ba:2016:VERA,
 author = {Ba, Mouhamadou Lamine and Berti-Equille, Laure and Shah, Kushal and Hammady, Hossam M.},
 title = {{VERA}: A Platform for Veracity Estimation over Web Data},
 booktitle = {Proceedings of the 25th International Conference Companion on World Wide Web},
 series = {WWW '16},
 year = {2016},
 isbn = {978-1-4503-4144-8},
 address = {Montr{\'e}al, Qu{\'e}bec, Canada},
 pages = {159--162},
 numpages = {4},
 acmid = {2890536},
 NOpublisher = {International World Wide Web Conferences Steering Committee},
 NOaddress = {Republic and Canton of Geneva, Switzerland},
 keywords = {data veracity, fact-checking, source trustworthiness, truth discovery},
}

@InProceedings{banerjee-han:2009:NAACLHLT09-Short,
  author    = {Banerjee, Protima  and  Han, Hyoil},
  title     = {Answer Credibility: A Language Modeling Approach to Answer Validation},
booktitle = {Proceedings of the Annual Conference of the North American Chapter of the Association for Computational Linguistics: Human Language Technologies},
  series = {NAACL-HLT~'09},
  NObooktitle = {Proceedings of HLT-NAACL, Companion Volume: Short Papers},
  NOmonth     = {June},
  year      = {2009},
  address   = {Boulder, CO, USA},
  NOpublisher = {Association for Computational Linguistics},
  pages     = {157--160}
}

@article{brill2001online,
  title={Online journalists embrace new marketing function},
  author={Brill, Ann M},
  journal={Newspaper Research Journal},
  volume={22},
  number={2},
  pages={28},
  year={2001},
  publisher={Newspaper Research Journal, Department of Journalism, University of Memphis}
}

@INPROCEEDINGS{Canini:2011,
author={Kevin R. Canini and Bongwon Suh and Peter L. Pirolli},
title={Finding Credible Information Sources in Social Networks Based on Content and Social Structure},
booktitle = {Proceedings of the IEEE International Conference on Privacy, Security, Risk, and Trust, and the IEEE International Conference on Social Computing},
series = {SocialCom/PASSAT~'11},
address = {Boston, MA, USA},
year={2011},
pages={1-8},
keywords={information management;social networking (online);user interfaces;content structure;credibility judgment;information credibility;information quality;information relevance;information source;social network user identification;social network user ranking;social structure;source credibility;Algorithm design and analysis;Analytical models;Appraisal;Tag clouds;Twitter;Visualization},
NOdoi={10.1109/PASSAT/SocialCom.2011.91},
month={Oct},}

@article{finberg2002digital,
  title={Digital journalism credibility study},
  author={Finberg, Howard and Stone, Martha L and Lynch, Diane},
  journal={Online News Association. Retrieved November},
  volume={3},
  pages={2003},
  year={2002}
}

@inproceedings{Jeon:2006:FPQ:1148170.1148212,
 author = {Jeon, Jiwoon and Croft, W. Bruce and Lee, Joon Ho and Park, Soyeon},
 title = {A Framework to Predict the Quality of Answers with Non-textual Features},
 booktitle = {Proceedings of the 29th Annual International ACM SIGIR Conference on Research and Development in Information Retrieval},
 series = {SIGIR '06},
 year = {2006},
 isbn = {1-59593-369-7},
 address = {Seattle, WA, USA},
 pages = {228--235},
 numpages = {8},
 acmid = {1148212},
 NOpublisher = {ACM},
 NOaddress = {New York, NY, USA},
 keywords = {document quality, information retrieval, language models, maximum entropy},
}

@inproceedings{Jurczyk:2007:DAQ:1321440.1321575,
 author = {Jurczyk, Pawel and Agichtein, Eugene},
 title = {Discovering Authorities in Question Answer Communities by Using Link Analysis},
 booktitle = {Proceedings of the 16th ACM Conference on Conference on Information and Knowledge Management},
 series = {CIKM '07},
 year = {2007},
 isbn = {978-1-59593-803-9},
 address = {Lisbon, Portugal},
 pages = {919--922},
 numpages = {4},
 acmid = {1321575},
 NOpublisher = {ACM},
 NOaddress = {New York, NY, USA},
 keywords = {link analysis, question-answer portals},
}

@inproceedings{lita2005qualitative,
 author = {Lita, Lucian Vlad and Schlaikjer, Andrew Hazen and Hong, WeiChang and Nyberg, Eric},
 title = {Qualitative Dimensions in Question Answering: Extending the Definitional {QA} Task},
 booktitle = {Proceedings of the 20th National Conference on Artificial Intelligence},
 series = {AAAI~'05},
 year = {2005},
 isbn = {1-57735-236-x},
 address = {Pittsburgh, PA, USA},
 pages = {1616--1617},
 numpages = {2},
 NOurl = {http://dl.acm.org/citation.cfm?id=1619566.1619583},
 NOacmid = {1619583},
 NOpublisher = {AAAI Press},
} 

@InProceedings{lukasik-cohn-bontcheva:2015:ACL-IJCNLP,
  author    = {Lukasik, Michal  and  Cohn, Trevor  and  Bontcheva, Kalina},
  title     = {Point Process Modelling of Rumour Dynamics in Social Media},
  booktitle = {Proceedings of the 53rd Annual Meeting of the Association for Computational Linguistics and the 7th International Joint Conference on Natural Language Processing},
  series    = {ACL-IJCNLP~'15},
  NOmonth     = {July},
  year      = {2015},
  address   = {Beijing, China},
  NOpublisher = {Association for Computational Linguistics},
  pages     = {518--523}
}

@inproceedings{lyon2001detecting,
  title={Detecting short passages of similar text in large document collections},
  author = {Lyon, Caroline and Malcolm, James and Dickerson, Bob},
  biburl = {http://www.bibsonomy.org/bibtex/2406fb7e1b89539289d30ecb501964362/saos},
  booktitle = {Proceedings of the Conference on Empirical Methods in Natural Language Processing},
  series = {EMNLP~'01},
  address = {Pittsburgh, PA, USA},
  interhash = {cb7da7544cc1edddc4441d7f7f595cb9},
  intrahash = {406fb7e1b89539289d30ecb501964362},
  keywords = {document_similarity},
  pages = {118--125},
  timestamp = {2014-06-25T15:30:23.000+0200},
  year = 2001
}

@inproceedings{RANLP2017:credibility:trolls,
  title={Do Not Trust the Trolls: Predicting Credibility in Community Question Answering Forums},
  author={Preslav Nakov and Tsvetomila Mihaylova and Llu\'is M\`arquez and Yashkumar Shiroya and Ivan Koychev},
  booktitle={Proceedings of the International Conference on Recent Advances in Natural Language Processing},
  series = {RANLP~'17},
  address = {Varna, Bulgaria},
  year={2017},
} 

@inproceedings{Ma:2015:DRU,
 author = {Ma, Jing and Gao, Wei and Wei, Zhongyu and Lu, Yueming and Wong, Kam-Fai},
 title = {Detect Rumors Using Time Series of Social Context Information on Microblogging Websites},
 booktitle = {Proceedings of the 24th ACM International on Conference on Information and Knowledge Management},
 series = {CIKM~'15},
 year = {2015},
 isbn = {978-1-4503-3794-6},
 address = {Melbourne, Australia},
 pages = {1751--1754},
 numpages = {4},
 acmid = {2806607},
 NOpublisher = {ACM},
 NOaddress = {New York, NY, USA},
 keywords = {rumor detection, social context, temporal, time series},
}

@inproceedings{Chen:2013:BIW:2492517.2492637,
 author = {Chen, Cheng and Wu, Kui and Srinivasan, Venkatesh and Zhang, Xudong},
 title = {Battling the {I}nternet {W}ater {A}rmy: detection of hidden paid posters},
 booktitle = {Proceedings of the 2013 IEEE/ACM International Conference on Advances in Social Networks Analysis and Mining},
 series = {ASONAM '13},
 year = {2013},
 isbn = {978-1-4503-2240-9},
 address = {Niagara, Canada},
 pages = {116--120},
 numpages = {5},
 NOdoi = {10.1145/2492517.2492637},
 acmid = {2492637},
} 

@inproceedings{ma2016detecting,
 author = {Ma, Jing and Gao, Wei and Mitra, Prasenjit and Kwon, Sejeong and Jansen, Bernard J. and Wong, Kam-Fai and Cha, Meeyoung},
 title = {Detecting Rumors from Microblogs with Recurrent Neural Networks},
 booktitle = {Proceedings of the 25th International Joint Conference on Artificial Intelligence},
 series = {IJCAI~'16},
 year = {2016},
 NOisbn = {978-1-57735-770-4},
 address = {New York, NY, USA},
 pages = {3818--3824},
 numpages = {7},
 NOurl = {http://dl.acm.org/citation.cfm?id=3061053.3061153},
 NOacmid = {3061153},
 NOpublisher = {AAAI Press},
} 

@article{Momeni:2015:SAR:2856149.2811282,
 author = {Momeni, Elaheh and Cardie, Claire and Diakopoulos, Nicholas},
 title = {A Survey on Assessment and Ranking Methodologies for User-Generated Content on the Web},
 journal = {ACM Comput. Surv.},
 issue_date = {February 2016},
 volume = {48},
 number = {3},
 month = dec,
 year = {2015},
 issn = {0360-0300},
 pages = {41:1--41:49},
 articleno = {41},
 numpages = {49},
 acmid = {2811282},
 publisher = {ACM},
 address = {New York, NY, USA},
 keywords = {User-generated content, adaptive, assessment, human centered, interactive, machine centered, ranking, social media},
}

@inproceedings{Morris:2012:TBU:2145204.2145274,
 author = {Morris, Meredith Ringel and Counts, Scott and Roseway, Asta and Hoff, Aaron and Schwarz, Julia},
 title = {Tweeting is Believing?: Understanding Microblog Credibility Perceptions},
 booktitle = {Proceedings of the ACM 2012 Conference on Computer Supported Cooperative Work},
 series = {CSCW '12},
 year = {2012},
 isbn = {978-1-4503-1086-4},
 location = {Seattle, WA, USA},
 pages = {441--450},
 numpages = {10},
 acmid = {2145274},
 publisher = {ACM},
 address = {New York, NY, USA},
 keywords = {credibility, microblogging, social search, twitter},
}

@inproceedings{mukherjee2015leveraging,
 author = {Mukherjee, Subhabrata and Weikum, Gerhard},
 title = {Leveraging Joint Interactions for Credibility Analysis in News Communities},
 booktitle = {Proceedings of the 24th ACM International on Conference on Information and Knowledge Management},
 series = {CIKM~'15},
 year = {2015},
 isbn = {978-1-4503-3794-6},
 address = {Melbourne, Australia},
 pages = {353--362},
 numpages = {10},
 NOurl = {http://doi.acm.org/10.1145/2806416.2806537},
 NOdoi = {10.1145/2806416.2806537},
 NOacmid = {2806537},
 NOpublisher = {ACM},
 NOaddress = {New York, NY, USA},
 keywords = {credibility, news community, probabilistic graphical models},
} 

@inproceedings{RANLP2017:clickbait,
  title={We Built a Fake News \& Click-bait Filter: What Happened Next Will Blow Your Mind!},
  author={Georgi Karadzhov and Pepa Gencheva and Preslav Nakov and Ivan Koychev},
  booktitle={Proceedings of the International Conference on Recent Advances in Natural Language Processing},
  series = {RANLP~'17},
  address = {Varna, Bulgaria},
  pages = {334--343},
  year={2017},
}

@inproceedings{SeminarUsers2017,
  author    = {Kareem Darwish and
               Dimitar Alexandrov and
               Preslav Nakov and
               Yelena Mejova},
  title     = {Seminar Users in the {A}rabic {T}witter Sphere},
  booktitle = {Proceedings of the 9th International Conference on Social Informatics},
  series    = {SocInfo~'17},
  address   = {Oxford, UK},
  pages     = {91--108},
  year      = {2017},
  NOurl       = {https://doi.org/10.1007/978-3-319-67217-5_7},
  NOdoi       = {10.1007/978-3-319-67217-5_7},
  NOtimestamp = {Wed, 06 Sep 2017 08:49:37 +0200},
  NObiburl    = {http://dblp.uni-trier.de/rec/bib/conf/socinfo/DarwishANM17},
  NObibsource = {dblp computer science bibliography, http://dblp.org}
}

@inproceedings{Pelleg:2016:CEI:2818048.2820022,
 author = {Pelleg, Dan and Rokhlenko, Oleg and Szpektor, Idan and Agichtein, Eugene and Guy, Ido},
 title = {When the Crowd is Not Enough: Improving User Experience with Social Media Through Automatic Quality Analysis},
 booktitle = {Proceedings of the 19th ACM Conference on Computer-Supported Cooperative Work \& Social Computing},
 series = {CSCW '16},
 year = {2016},
 isbn = {978-1-4503-3592-8},
 address = {San Francisco, CA, USA},
 pages = {1080--1090},
 numpages = {11},
 acmid = {2820022},
 NOpublisher = {ACM},
 NOaddress = {New York, NY, USA},
 keywords = {A/B testing, Automatic quality evaluation, Quantitative analysis, User engagement},
}

@article{PlosONE:2016,
  Author = {Zubiaga, Arkaitz AND Liakata, Maria AND Procter, Rob AND Wong Sak Hoi, Geraldine AND Tolmie, Peter},
  Journal = {PLoS ONE},
  Month = {03},
  Number = {3},
  Pages = {1-29},
  Publisher = {Public Library of Science},
  Title = {Analysing How People Orient to and Spread Rumours in Social Media by Looking at Conversational Threads},
  Volume = {11},
  Year = {2016}}

@inproceedings{popat2016credibility,
 author = {Popat, Kashyap and Mukherjee, Subhabrata and Str\"{o}tgen, Jannik and Weikum, Gerhard},
 title = {Credibility Assessment of Textual Claims on the Web},
 booktitle = {Proceedings of the 25th ACM International on Conference on Information and Knowledge Management},
 series = {CIKM~'16},
 year = {2016},
 isbn = {978-1-4503-4073-1},
 address = {Indianapolis, IN, USA},
 pages = {2173--2178},
 numpages = {6},
 NOurl = {http://doi.acm.org/10.1145/2983323.2983661},
 NOdoi = {10.1145/2983323.2983661},
 acmid = {2983661},
 NOpublisher = {ACM},
 NOaddress = {New York, NY, USA},
 keywords = {credibility analysis, rumor and hoax detection, text mining},
} 

@inproceedings{potthast2013overview,
  title={Overview of the 5th International Competition on Plagiarism Detection},
  author={Potthast, Martin and Hagen, Matthias and Gollub, Tim and Tippmann, Martin and Kiesel, Johannes and Rosso, Paolo and Stamatatos, Efstathios and Stein, Benno},
  booktitle={Proceedings of the Conference on Multilingual and Multimodal Information Access Evaluation},
  series = {CLEF~'13},
  pages={301--331},
  year={2013},
  address = {Valencia, Spain},
  NOorganization={CELCT}
}

@inproceedings{Su-EtAl:2010:PACLIC2010,
author    = {Qi Su and Helen Kai-Yun Chen and Chu-Ren Huang},
title     = {Incorporate Credibility into Context for the Best Social Media Answers},
booktitle = {Proceedings of the 24th Pacific Asia Conference on Language, Information and Computation},
series    = {PACLIC~'10},
year      = {2010},
NOmonth     = {November},
address   = {Sendai, Japan},
NOpublisher = {Institute of Digital Enhancement of Cognitive Processing, Waseda University},
pages     = {535--541},
NOurl       = {http://www.aclweb.org/anthology/Y10-1062}
}

@inproceedings{Zaharia:2010:SCC:1863103.1863113,
 author = {Zaharia, Matei and Chowdhury, Mosharaf and Franklin, Michael J. and Shenker, Scott and Stoica, Ion},
 title = {Spark: Cluster Computing with Working Sets},
 booktitle = {Proceedings of the 2nd USENIX Conference on Hot Topics in Cloud Computing},
 series = {HotCloud~'10},
 year = {2010},
 address = {Boston, MA, USA},
 pages = {10--10},
 numpages = {1},
 acmid = {1863113},
 NOpublisher = {USENIX Association},
 NOaddress = {Berkeley, CA, USA},
}

@inproceedings{RANLP2017:debates,
  title={A Context-Aware Approach for Detecting Worth-Checking Claims in Political Debates},
  author={Pepa Gencheva and Preslav Nakov and Llu\'{i}s M\`{a}rquez and Alberto Barr\'on-Cede{\~n}o and Ivan Koychev},
  booktitle={Proceedings of the International Conference on Recent Advances in Natural Language Processing},
  series = {RANLP~'17},
  address = {Varna, Bulgaria},
  year={2017},
}

@article{zubiaga2015analysing,
  title={Analysing How People Orient to and Spread Rumours in Social Media by Looking at Conversational Threads},
  author={Zubiaga, Arkaitz and Hoi, Geraldine Wong Sak and Liakata, Maria and Procter, Rob and Tolmie, Peter},
  journal={arXiv preprint arXiv:1511.07487},
  year={2015}
}



@ARTICLE{bowman:reasoning,
    author = {Bowman, Mic and Debray, Saumya K. and Peterson, Larry L.},
    title = {Reasoning About Naming Systems},
    journal = {ACM Trans. Program. Lang. Syst.},
    volume = {15},
    number = {5},
    pages = {795-825},
    month = {November},
    year = {1993},
    doi = {10.1145/161468.161471},
}

@ARTICLE{braams:babel,
    author = {Braams, Johannes},
    title = {Babel, a Multilingual Style-Option System for Use with LaTeX's Standard Document Styles},
    journal = {TUGboat},
    volume = {12},
    number = {2},
    pages = {291-301},
    month = {June},
    year = {1991},
}

@INPROCEEDINGS{clark:pct,
	AUTHOR = "Malcolm Clark",
	TITLE = "Post Congress Tristesse",
	BOOKTITLE = "TeX90 Conference Proceedings",
	PAGES = "84-89",
	ORGANIZATION = "TeX Users Group",
	MONTH = "March", 
	YEAR = {1991}	}


@Inproceedings{Hardalov2016,
author="Hardalov, Momchil
and Koychev, Ivan
and Nakov, Preslav",
NOeditor="Dichev, Christo and Agre, Gennady",
title="In Search of Credible News",
bookTitle="Proceedings of the 17th International Conference on Artificial Intelligence: Methodology, Systems, and Applications",
series = {AIMSA~'16},
address = {Varna, Bulgaria},
year="2016",
NOpublisher="Springer International Publishing",
pages="172--180",
isbn="978-3-319-44748-3",
NOdoi="10.1007/978-3-319-44748-3_17",
NOurl="https://doi.org/10.1007/978-3-319-44748-3_17"
}

@ARTICLE{herlihy:methodology,
    author = {Herlihy, Maurice},
    title = {A Methodology for Implementing Highly Concurrent Data Objects},
    journal = {ACM Trans. Program. Lang. Syst.},
    volume = {15},
    number = {5},
    pages = {745-770},
    month = {November},
    year = {1993},
    doi = {10.1145/161468.161469},
}

@BOOK{Lamport:LaTeX,
	AUTHOR = "Leslie Lamport",
	TITLE = "LaTeX User's Guide and Document Reference Manual",
	PUBLISHER = "Addison-Wesley Publishing Company",
	ADDRESS = "Reading, Massachusetts",
	YEAR = "1986"	}

@BOOK{salas:calculus,
	AUTHOR = "S.L. Salas and Einar Hille",
	TITLE = "Calculus: One and Several Variable",
	PUBLISHER = "John Wiley and Sons",
	ADDRESS = "New York",
	YEAR = "1978"	}

@MANUAL{Fear05,
  title = 	 {Publication quality tables in {\LaTeX}},
  author =	 {Simon Fear},
  month =	 {April},
  year =	 2005,
  note =	 {\url{http://www.ctan.org/pkg/booktabs}}
}

@InProceedings{mihaylov-nakov:2016:P16-2,
  author    = {Mihaylov, Todor  and  Nakov, Preslav},
  title     = {{H}unting for troll comments in news community forums},
  booktitle = {Proceedings of the 54th Annual Meeting of the Association for Computational Linguistics},
  series    = {ACL~'16},
  NOmonth     = {August},
  year      = {2016},
  address   = {Berlin, Germany},
  pages     = {399--405},
  NOurl       = {http://anthology.aclweb.org/P16-2065}
}

@article{Liu:2016:SGD:2872563.2872566,
 author = {Liu, Dong and Wu, Quanyuan and Han, Weihong and Zhou, Bin},
 title = {{S}ockpuppet gang detection on social media sites},
 journal = {Front. Comput. Sci.},
 issue_date = {February  2016},
 volume = {10},
 number = {1},
 month = feb,
 year = {2016},
 issn = {2095-2228},
 pages = {124--135},
 numpages = {12},
 NOurl = {http://dx.doi.org/10.1007/s11704-015-4287-7},
 NOdoi = {10.1007/s11704-015-4287-7},
 acmid = {2872566},
 publisher = {Springer-Verlag New York, Inc.},
 address = {Secaucus, NJ, USA},
 keywords = {sentiment orientation, social media site, sockpuppet gang detection, user behavior feature},
} 


@inproceedings{Mihaylov2015FindingOM,
  author    = {Mihaylov, Todor  and  Georgiev, Georgi  and  Nakov, Preslav},
  title     = {{F}inding opinion manipulation trolls in news community forums},
  booktitle = {Proceedings of the Nineteenth Conference on Computational Natural Language Learning},
  series = {CoNLL~'15},
  NOmonth     = {July},
  year      = {2015},
  address   = {Beijing, China},
  pages     = {310--314},
  NOurl       = {http://www.aclweb.org/anthology/K15-1032}
}


@inproceedings{Mihaylov2015ExposingPO,
  author    = {Mihaylov, Todor  and  Koychev, Ivan  and  Georgiev, Georgi  and  Nakov, Preslav},
  title     = {{E}xposing paid opinion manipulation trolls},
  booktitle = {Proceedings of the International Conference Recent Advances in Natural Language Processing},
  series = {RANLP~'15},
  NOmonth     = {September},
  year      = {2015},
  address   = {Hissar, Bulgaria},
  NOpublisher = {INCOMA Ltd. Shoumen, BULGARIA},
  pages     = {443--450},
  NOurl       = {http://www.aclweb.org/anthology/R15-1058}
}

@Manual{Amsthm15,
  title = 	 {Using the amsthm Package},
  organization = {American Mathematical Society},
  month =	 {April},
  year =	 2015,
  note =	 {\url{http://www.ctan.org/pkg/amsthm}}
}

@inproceedings{baly2018predicting,
  title       = {Predicting Factuality of Reporting and Bias of News Media Sources},
  author      = {Baly, Ramy and Karadzhov, Georgi and Alexandrov, Dimitar and Glass, James and Nakov, Preslav},
  booktitle   = {Proceedings of the Conference on Empirical Methods in Natural Language Processing},
  NOmonth     = {October},
  year        = {2018},
  series      = {EMNLP~'18},
  NOpublisher = {Association for Computational Linguistics},
  address = {Brussels, Belgium},
  pages={3528--3539}
}

@InProceedings{pennington-socher-manning:2014:EMNLP2014,
  author    = {Pennington, Jeffrey  and  Socher, Richard  and  Manning, Christopher},
  title     = {{GloVe}: Global Vectors for Word Representation},
  booktitle = {Proceedings of the Conference on Empirical Methods in Natural Language Processing},
  series = {EMNLP~'14},
  NOmonth     = {October},
  year      = {2014},
  address   = {Doha, Qatar},
  NOpublisher = {Association for Computational Linguistics},
  pages     = {1532--1543},
  NOurl       = {http://www.aclweb.org/anthology/D14-1162}
}

@InProceedings{nakov-EtAl:2016:SemEval,
  author    = {Nakov, Preslav  and  M\`{a}rquez, Llu\'{i}s  and  Moschitti, Alessandro  and  Magdy, Walid  and  Mubarak, Hamdy  and  Freihat, abed Alhakim  and  Glass, Jim  and  Randeree, Bilal},
  title     = {{SemEval}-2016 Task 3: Community Question Answering},
  booktitle = {Proceedings of the 10th International Workshop on Semantic Evaluation},
  series    = {SemEval~'16},
  NOmonth     = {June},
  year      = {2016},
  address   = {San Diego, CA, USA},
  NOpublisher = {Association for Computational Linguistics},
  pages     = {525--545},
  NOurl       = {http://www.aclweb.org/anthology/S16-1083}
}

@inproceedings{RANLP2017:factchecking:external,
  title={Fully Automated Fact Checking Using External Sources},
  author={Georgi Karadzhov and Preslav Nakov and Llu\'{i}s M\`{a}rquez and Alberto Barr\'on-Cede{\~n}o and Ivan Koychev},
  booktitle={Proceedings of the International Conference on Recent Advances in Natural Language Processing},
  series = {RANLP~'17},
  address = {Varna, Bulgaria},
  year={2017},
  pages     = {344--353},
  NOURL="https://arxiv.org/abs/1710.00341"
}

@InProceedings{nakov-EtAl:2017:SemEval,
  author    = {Nakov, Preslav  and  Hoogeveen, Doris  and  M\`{a}rquez, Llu\'{i}s  and  Moschitti, Alessandro  and  Mubarak, Hamdy  and  Baldwin, Timothy  and  Verspoor, Karin},
  title     = {{SemEval}-2017 Task 3: Community Question Answering},
  booktitle = {Proceedings of the 11th International Workshop on Semantic Evaluation},
  series    = {SemEval~'17},
  NOmonth     = {August},
  year      = {2017},
  address   = {Vancouver, Canada},
  NOpublisher = {Association for Computational Linguistics},
  pages     = {27--48},
  NOurl       = {http://www.aclweb.org/anthology/S17-2003}
}

@INPROCEEDINGS{Kwon:2013, 
author={Sejeong Kwon and Meeyoung Cha and Kyomin Jung and Wei Chen and Yajun Wang}, 
title={Prominent Features of Rumor Propagation in Online Social Media}, 
booktitle={Proceedings of the 13th IEEE International Conference on Data Mining}, 
series = {ICDM~'13},
year={2013}, 
pages={1103-1108}, 
address = {Dallas, TX, USA},
keywords={social networking (online);time series;daily shock cycles;external shock cycles;linguistic differences;online social media;online social networks;periodic time series model;rumor classification;rumor propagation;structural differences;temporal characteristics;Adaptation models;Electric shock;Mathematical model;Pragmatics;Psychology;Time series analysis;Twitter;Diffusion Network;Rumor;Sentiment Analysis;Social Media;Time Series}, 
NOdoi={10.1109/ICDM.2013.61}, 
ISSN={1550-4786}, 
month={Dec},}

@inproceedings{Maity:2017:DSS:3022198.3026360,
 author = {Maity, Suman Kalyan and Chakraborty, Aishik and Goyal, Pawan and Mukherjee, Animesh},
 title = {{D}etection of sockpuppets in social media},
 booktitle = {Proceedings of the ACM Conference on Computer Supported Cooperative Work and Social Computing},
 series = {CSCW~'17},
 year = {2017},
 isbn = {978-1-4503-4688-7},
 address = {Portland, OR, USA},
 pages = {243--246},
 numpages = {4},
 NOurl = {http://doi.acm.org/10.1145/3022198.3026360},
 NOdoi = {10.1145/3022198.3026360},
 acmid = {3026360},
 keywords = {classification, social media, sockpuppet},
} 

@inproceedings{Kumar:2017:AMS:3038912.3052677,
 author = {Kumar, Srijan and Cheng, Justin and Leskovec, Jure and Subrahmanian, V.S.},
 title = {{A}n army of me: sockpuppets in online discussion communities},
 booktitle = {Proceedings of the 26th International Conference on World Wide Web},
 series = {WWW '17},
 year = {2017},
 isbn = {978-1-4503-4913-0},
 address = {Perth, Australia},
 pages = {857--866},
 numpages = {10},
 NOurl = {https://doi.org/10.1145/3038912.3052677},
 NOdoi = {10.1145/3038912.3052677},
 NOacmid = {3052677},
 NOpublisher = {International World Wide Web Conferences Steering Committee},
 NOaddress = {Republic and Canton of Geneva, Switzerland},
 keywords = {antisocial behavior, malicious users, multiple account use},
} 


@article{journals/intr/CastilloMP13,
  author = {Castillo, Carlos and Mendoza, Marcelo and Poblete, Barbara},
  journal = {Internet Research},
  keywords = {dblp},
  number = 5,
  pages = {560-588},
  title = {Predicting information credibility in time-sensitive social media.},
  NOurl = {http://dblp.uni-trier.de/db/journals/intr/intr23.html#CastilloMP13},
  volume = 23,
  year = 2013
}

@phdthesis{phdthesis:graves,
  author       = {Graves, Lucas}, 
  title        = {Deciding What's True: Fact-Checking Journalism and the New Ecology of News},
  school       = {Columbia University},
  year         = 2013,
}

@Article{Zubiaga2014,
author="Zubiaga, Arkaitz
and Ji, Heng",
title="Tweet, but verify: epistemic study of information verification on {T}witter",
journal="Social Network Analysis and Mining",
year="2014",
volume="4",
number="1",
pages="1--12",
issn="1869-5469",
NOdoi="10.1007/s13278-014-0163-y",
NOurl="http://dx.doi.org/10.1007/s13278-014-0163-y"
}

@article{Papadopoulos:2016:OSI,
 author = {Papadopoulos, Symeon and Bontcheva, Kalina and Jaho, Eva and Lupu, Mihai and Castillo, Carlos},
 title = {Overview of the Special Issue on Trust and Veracity of Information in Social Media},
 journal = {ACM Trans. Inf. Syst.},
 issue_date = {May 2016},
 volume = {34},
 number = {3},
 month = apr,
 year = {2016},
 issn = {1046-8188},
 pages = {14:1--14:5},
 articleno = {14},
 numpages = {5},
 NOurl = {http://doi.acm.org/10.1145/2870630},
 NOdoi = {10.1145/2870630},
 acmid = {2870630},
 NOpublisher = {ACM},
 address = {New York, NY, USA},
 keywords = {Information veracity, fake content, news verification, rumour propagation, social media},
} 


@InProceedings{derczynski-EtAl:2017:SemEval,
  author    = {Derczynski, Leon  and  Bontcheva, Kalina  and  Liakata, Maria  and  Procter, Rob  and  Wong Sak Hoi, Geraldine  and  Zubiaga, Arkaitz},
  title     = {{SemEval-2017 Task 8: RumourEval}: Determining rumour veracity and support for rumours},
  booktitle = {Proceedings of the 11th International Workshop on Semantic Evaluation},
  series = {SemEval~'17},
  NOmonth     = {August},
  year      = {2017},
  address   = {Vancouver, Canada},
  NOpublisher = {Association for Computational Linguistics},
  pages     = {60--67},
  NOurl       = {http://www.aclweb.org/anthology/S17-2006}
}

@article{Zanzoto-RTE,
author = {Zanzotto, Fabio Massimo and Pennacchiotti, Marco and Moschitti, Alessandro},
year = {2009},
month = {10},
pages = {551-582},
title = {A machine learning approach to textual entailment recognition},
volume = {15},
journal = {Natural Language Engineering}
}

@inproceedings{Wei:2015:FTS:2808797.2809316,
 author = {Wei, Wei and Joseph, Kenneth and Liu, Huan and Carley, Kathleen M.},
 title = {{T}he fragility of {T}witter social networks against suspended users},
 booktitle = {Proceedings of the 2015 IEEE/ACM International Conference on Advances in Social Networks Analysis and Mining 2015},
 series = {ASONAM '15},
 year = {2015},
 isbn = {978-1-4503-3854-7},
 address = {Paris, France},
 pages = {9--16},
 numpages = {8},
 acmid = {2809316},
} 

@INPROCEEDINGS{clickbait:2016, 
author={Chakraborty, Abhijnan and Paranjape, Bhargavi and Kakarla, Kakarla and Ganguly, Niloy}, 
booktitle={Proceedings of the 2016 IEEE/ACM International Conference on Advances in Social Networks Analysis and Mining}, 
title={Stop Clickbait: Detecting and preventing clickbaits in online news media}, 
series={ASONAM~'16},
year={2016}, 
volume={}, 
number={}, 
pages={9-16}, 
address = {San Francisco, CA, USA},
keywords={online front-ends;social networking (online);browser extension;clickbait;online news media;reader attention;Browsers;Facebook;Internet;Media;Syntactics;Web pages}, 
NOdoi={10.1109/ASONAM.2016.7752207}, 
ISSN={}, 
NOmonth={Aug}
}

@inproceedings{gilbert2014vader,
  title={{VADER}: A parsimonious rule-based model for sentiment analysis of social media text},
  author={Hutto, Clayton and Gilbert, Eric},
  booktitle={Proceedings of the 8th International Conference on Weblogs and Social Media},
  series = {ICWSM~'14},
  address = {Ann Arbor, MI, USA},
  year={2014}
}

@inproceedings{recasens2013linguistic,
  title={Linguistic models for analyzing and detecting biased language},
  author={Recasens, Marta and Danescu-Niculescu-Mizil, Cristian and Jurafsky, Dan},
  booktitle={Proceedings of the 51st Annual Meeting of the Association for Computational Linguistics},
  series = {ACL~'13},
  pages={1650--1659},
  address = {Sofia, Bulgaria},
  year={2013}
}

@article{mitchell2013geography,
  title={The geography of happiness: Connecting {T}witter sentiment and expression, demographics, and objective characteristics of place},
  author={Mitchell, Lewis and Frank, Morgan R and Harris, Kameron Decker and Dodds, Peter Sheridan and Danforth, Christopher M},
  journal={PloS one},
  volume={8},
  number={5},
  pages={e64417},
  year={2013},
  publisher={Public Library of Science}
}

@inproceedings{horne2017identifying,
  title={Identifying the social signals that drive online discussions: A case study of {Reddit} communities},
  author={Horne, Benjamin and Adali, Sibel and Sikdar, Sujoy},
  booktitle={Proceedings of the 26th IEEE International Conference on Computer Communication and Networks},
  series = {ICCCN~'17},
  address = {Vancouver, Canada},
  pages={1--9},
  year={2017},
  NOorganization={IEEE}
}

@article{graham2009liberals,
  title={Liberals and conservatives rely on different sets of moral foundations.},
  author={Graham, Jesse and Haidt, Jonathan and Nosek, Brian A},
  journal={Journal of personality and social psychology},
  volume={96},
  number={5},
  pages={1029},
  year={2009},
  publisher={US: American Psychological Association}
}

@article{lin2017acquiring,
  title={Acquiring background knowledge to improve moral value prediction},
  author={Lin, Ying and Hoover, Joe and Dehghani, Morteza and Mooijman, Marlon and Ji, Heng},
  journal={arXiv preprint arXiv:1709.05467},
  year={2017}
}


@article{basnet2014learning,
  title={Learning to detect phishing {URLs}},
  author={Basnet, Ram B and Sung, Andrew H and Liu, Quingzhong},
  journal={International Journal of Research in Engineering and Technology},
  volume={3},
  number={6},
  pages={11--24},
  year={2014}
}

@InProceedings{clef2018checkthat:overall,
    author    = {Nakov, Preslav  and  Barr\'{o}n-Cede\~{n}o, Alberto and Elsayed, Tamer and Suwaileh, Reem and M\`{a}rquez, Llu\'{i}s and Zaghouani, Wajdi and Atanasova, Pepa and Kyuchukov, Spas and Da San Martino, Giovanni},
    title     = {Overview of the {CLEF-2018 CheckThat!} Lab on Automatic Identification and Verification of Political Claims},
    booktitle = {Proceedings of the Ninth International Conference of the CLEF Association: Experimental IR Meets Multilinguality, Multimodality, and Interaction},
    series    = {Lecture Notes in Computer Science},
    publisher = {Springer},
    NOeditor    = {Patrice Bellot, Chiraz Trabelsi, Josiane Mothe, Fionn Murtagh, Jian Yun Nie, Laure Soulier, Eric Sanjuan, Linda Cappellato, Nicola Ferro},
    address   = {Avignon, France},
    NOmonth     = {September},
    year      = {2018},
    pages = {372--387}
}

@InProceedings{clef2018checkthat:task1,
    author    = {Atanasova, Pepa and  M\`{a}rquez, Llu\'{i}s  and Barr\'{o}n-Cede\~{n}o, Alberto  and Elsayed, Tamer and Suwaileh, Reem and Zaghouani, Wajdi and Kyuchukov, Spas and Da San Martino, Giovanni and Nakov, Preslav},
    title     = {Overview of the {CLEF-2018 CheckThat!} Lab on Automatic Identification and Verification of Political Claims, Task 1: Check-Worthiness},
    booktitle = {CLEF 2018 Working Notes. Working Notes of CLEF 2018 - Conference and Labs of the Evaluation Forum},
    series    = {{CEUR} Workshop Proceedings},
    publisher = {CEUR-WS.org},
    NOeditor    = {Cappellato, Linda and Ferro, Nicola and Nie, Jian-Yun and Soulier, Laure},
    address   = {Avignon, France},
    NOmonth     = {September},
    year      = {2018},
}


@InProceedings{clef2018checkthat:task2,
    author    = {Barr\'{o}n-Cede\~{n}o, Alberto and Elsayed, Tamer and Suwaileh, Reem and M\`{a}rquez, Llu\'{i}s  and Atanasova, Pepa and Zaghouani, Wajdi and Kyuchukov, Spas and Da San Martino, Giovanni and Nakov, Preslav},
    title     = {Overview of the {CLEF-2018 CheckThat!} Lab on Automatic Identification and Verification of Political Claims, Task 2: Factuality},
    booktitle = {CLEF 2018 Working Notes. Working Notes of CLEF 2018 - Conference and Labs of the Evaluation Forum},
    series    = {{CEUR} Workshop Proceedings},
    publisher = {CEUR-WS.org},
    NOeditor    = {Cappellato, Linda and Ferro, Nicola and Nie, Jian-Yun and Soulier, Laure},
    address   = {Avignon, France},
    NOmonth     = {September},
    year      = {2018},
}

@InProceedings{thorne-EtAl:2018:N18-1,
  author    = {Thorne, James  and  Vlachos, Andreas  and  Christodoulopoulos, Christos  and  Mittal, Arpit},
  title     = {{FEVER}: a Large-scale Dataset for Fact Extraction and {VERification}},
  booktitle = {Proceedings of the 2018 Conference of the North American Chapter of the Association for Computational Linguistics: Human Language Technologies},
  series = {NAACL-HLT~'18},
  NOmonth     = {June},
  year      = {2018},
  address   = {New Orleans, LA, USA},
  NOpublisher = {Association for Computational Linguistics},
  pages     = {809--819},
  NOurl       = {http://www.aclweb.org/anthology/N18-1074}
}

@InProceedings{thorne-vlachos:2018:C18-1,
  author    = {Thorne, James  and  Vlachos, Andreas},
  title     = {Automated Fact Checking: Task Formulations, Methods and Future Directions},
  booktitle = {Proceedings of the 27th International Conference on Computational Linguistics},
  series = {COLING~'18},
  NOmonth     = {August},
  year      = {2018},
  address   = {Santa Fe, NM, USA},
  NOpublisher = {Association for Computational Linguistics},
  pages     = {3346--3359},
  NOurl       = {http://www.aclweb.org/anthology/C18-1283}
}

@InProceedings{dungs-EtAl:2018:C18-1,
  author    = {Dungs, Sebastian  and  Aker, Ahmet  and  Fuhr, Norbert  and  Bontcheva, Kalina},
  title     = {Can Rumour Stance Alone Predict Veracity?},
  booktitle = {Proceedings of the 27th International Conference on Computational Linguistics},
  series    = {COLING~'18},
  NOmonth     = {August},
  year      = {2018},
  address   = {Santa Fe, NM, USA},
  NOpublisher = {Association for Computational Linguistics},
  pages     = {3360--3370},
  NOurl       = {http://www.aclweb.org/anthology/C18-1284}
}

@InProceedings{desarkar-yang-mukherjee:2018:C18-1,
  author    = {De Sarkar, Sohan  and  Yang, Fan  and  Mukherjee, Arjun},
  title     = {Attending Sentences to detect Satirical Fake News},
  booktitle = {Proceedings of the 27th International Conference on Computational Linguistics},
  series    = {COLING~'18},
  NOmonth     = {August},
  year      = {2018},
  address   = {Santa Fe, NM, USA},
  NOpublisher = {Association for Computational Linguistics},
  pages     = {3371--3380},
  NOurl       = {http://www.aclweb.org/anthology/C18-1285}
}

@InProceedings{prezrosas-EtAl:2018:C18-1,
  author    = {P\'{e}rez-Rosas, Ver\'{o}nica  and  Kleinberg, Bennett  and  Lefevre, Alexandra  and  Mihalcea, Rada},
  title     = {Automatic Detection of Fake News},
  booktitle = {Proceedings of the 27th International Conference on Computational Linguistics},
  series    = {COLING~'18},
  NOmonth     = {August},
  year      = {2018},
  address   = {Santa Fe, NM, USA},
  NOpublisher = {Association for Computational Linguistics},
  pages     = {3391--3401},
  NOurl       = {http://www.aclweb.org/anthology/C18-1287}
}

@InProceedings{kochkina-liakata-zubiaga:2018:C18-1,
  author    = {Kochkina, Elena  and  Liakata, Maria  and  Zubiaga, Arkaitz},
  title     = {All-in-one: Multi-task Learning for Rumour Verification},
  booktitle = {Proceedings of the 27th International Conference on Computational Linguistics},
  series    = {COLING~'18},
  NOmonth     = {August},
  year      = {2018},
  address   = {Santa Fe, NM, USA},
  NOpublisher = {Association for Computational Linguistics},
  pages     = {3402--3413},
  NOurl       = {http://www.aclweb.org/anthology/C18-1288}
}

@article{InternetResearchJournal:2018,
  author    = {Mihaylov, Todor and Mihaylova, Tsvetomila and Nakov, Preslav and M\`{a}rquez, Llu\'{i}s and Georgiev, Georgi and Koychev, Ivan},
  title     = {The Dark Side of News Community Forums: Opinion Manipulation Trolls},
  journal   = {Internet Research},
  year      = {2018},
  volume    = {28},
  number    = {5},
  pages     = {1292--1312},
}

@InProceedings{AAAI2018:factchecking,
  author    = {Tsvetomila Mihaylova and Preslav Nakov and Llu\'{i}s M\`{a}rquez and Alberto Barr\'on-Cede{\~n}o and Mitra Mohtarami and Georgi Karadjov and James Glass},
  title     = {Fact Checking in Community Forums},
  booktitle = {Proceedings of the Thirty-Second AAAI Conference on Artificial Intelligence},
  series    = {AAAI~'18},
  year      = {2018},
  address   = {New Orleans, LA, USA},
  pages     = {879--886},
  month     = {February},
  noURL     = "https://www.aaai.org/ocs/index.php/AAAI/AAAI18/paper/viewFile/16780/16082"
}


@InProceedings{Pan:KG:2018,
  author    = {Jeff Z. Pan and Siyana Pavlova and Chenxi Li and Ningxi Li and Yangmei Li and Jinshuo Liu},
  title     = {Content based Fake News Detection Using Knowledge Graphs},
  booktitle = {Proceedings of the International Semantic Web Conference},
  series    = {ISWC~'18},
  year      = {2018},
  address   = {Monterey, CA, USA},
}

@InProceedings{baly-EtAl:2018:N18-2,
  author    = {Baly, Ramy  and  Mohtarami, Mitra  and  Glass, James  and  M\`{a}rquez, Llu\'{i}s  and  Moschitti, Alessandro  and  Nakov, Preslav},
  title     = {Integrating Stance Detection and Fact Checking in a Unified Corpus},
  booktitle = {Proceedings of the 2018 Conference of the North American Chapter of the Association for Computational Linguistics: Human Language Technologies},
  series    = {NAACL-HLT~'18},
  NOmonth     = {June},
  year      = {2018},
  address   = {New Orleans, LA, USA},
  NOpublisher = {Association for Computational Linguistics},
  pages     = {21--27},
  NOurl       = {http://www.aclweb.org/anthology/N18-2004}
}

@article{Pitoura:2018:MBO:3186549.3186553,
 author = {Pitoura, Evaggelia and Tsaparas, Panayiotis and Flouris, Giorgos and Fundulaki, Irini and Papadakos, Panagiotis and Abiteboul, Serge and Weikum, Gerhard},
 title = {On Measuring Bias in Online Information},
 journal = {SIGMOD Rec.},
 issue_date = {December 2017},
 volume = {46},
 number = {4},
 month = feb,
 year = {2018},
 issn = {0163-5808},
 pages = {16--21},
 numpages = {6},
 NOurl = {http://doi.acm.org/10.1145/3186549.3186553},
 NOdoi = {10.1145/3186549.3186553},
 acmid = {3186553},
 publisher = {ACM},
 NOaddress = {New York, NY, USA},
}

@article{riedel2017simple,
	author={Riedel, Benjamin and Augenstein, Isabelle and Spithourakis, Georgios P and Riedel, Sebastian},
    title={A Simple but Tough-to-beat Baseline for the {Fake News Challenge} Stance Detection Task},
    journal={ArXiv:1707.03264},
    archivePrefix={arXiv},
    eprint={1707.03264},
    primaryClass={cs.CL},
    keywords={Computer Science - Computation and Language},
    adsurl={http://adsabs.harvard.edu/abs/2017arXiv170703264R},
    adsnote={Provided by the SAO/NASA Astrophysics Data System},
    NOmonth={July},
    year={2017}
}

@InProceedings{NAACL2018:stance,
  author    = {Mitra Mohtarami and Ramy Baly and James Glass and Preslav Nakov and Llu\'{i}s M\`{a}rquez and Alessandro Moschitti},
  title     = {Automatic Stance Detection Using End-to-End Memory Networks},
  booktitle = {Proceedings of the 16th Annual Conference of the North American Chapter of the Association for Computational Linguistics: Human Language Technologies},
  series    = {NAACL-HLT~'18},
  year      = {2018},
  address   = {New Orleans, LA, USA},
  NOmonth     = {June},
  pages = {767--776},
  noURL     = "http://aclweb.org/anthology/N18-1070"
}

@InProceedings{hanselowski-EtAl:2018:C18-1,
  author    = {Hanselowski, Andreas  and  PVS, Avinesh  and  Schiller, Benjamin  and  Caspelherr, Felix  and  Chaudhuri, Debanjan  and  Meyer, Christian M.  and  Gurevych, Iryna},
  title     = {A Retrospective Analysis of the Fake News Challenge Stance-Detection Task},
  booktitle = {Proceedings of the 27th International Conference on Computational Linguistics},
  series    = {COLING~'18},
  NOmonth     = {August},
  year      = {2018},
  address   = {Santa Fe, NM, USA},
  NOpublisher = {Association for Computational Linguistics},
  pages     = {1859--1874},
  NOurl       = {http://www.aclweb.org/anthology/C18-1158}
}

@InProceedings{DarwishMZ17,
	author={Darwish, Kareem and Magdy, Walid and Zanouda, Tahar},
    title={Improved Stance Prediction in a User Similarity Feature Space},
    booktitle={Proceedings of the Conference on Advances in Social Networks Analysis and Mining},
    series = {ASONAM~'17},
    pages={145--148},
    address={Sydney, Australia},
    year={2017}
}

@inproceedings{Popat:2018:CCL:3184558.3186967,
 author = {Popat, Kashyap and Mukherjee, Subhabrata and Str\"{o}tgen, Jannik and Weikum, Gerhard},
 title = {{CredEye}: A Credibility Lens for Analyzing and Explaining Misinformation},
 booktitle = {Proceedings of The Web Conference 2018},
 series = {WWW~'18},
 year = {2018},
 NOisbn = {978-1-4503-5640-4},
 address = {Lyon, France},
 pages = {155--158},
 numpages = {4},
 NOurl = {https://doi.org/10.1145/3184558.3186967},
 NOdoi = {10.1145/3184558.3186967},
 NOacmid = {3186967},
 NOpublisher = {International World Wide Web Conferences Steering Committee},
 NOaddress = {Republic and Canton of Geneva, Switzerland},
 keywords = {credibility analysis, fact checking, interpretable learning},
} 

@InProceedings{thorne-EtAl:2017:NLPmJ,
  author    = {Thorne, James  and  Chen, Mingjie  and  Myrianthous, Giorgos  and  Pu, Jiashu  and  Wang, Xiaoxuan  and  Vlachos, Andreas},
  title     = {Fake news stance detection using stacked ensemble of classifiers},
  booktitle = {Proceedings of the EMNLP Workshop on Natural Language Processing meets Journalism},
  NOmonth   = {September},
  year      = {2017},
  address   = {Copenhagen, Denmark},
  NOpublisher = {Association for Computational Linguistics},
  pages     = {80--83},
  NOurl     = {http://www.aclweb.org/anthology/W17-4214}
}

@article{ZubiagaKLPLBCA18,
  author    = {Zubiaga, Arkaitz and Kochkina, Elena and Liakata, Maria and Procter, Rob and Lukasik, Michal and Bontcheva, Kalina and Cohn, Trevor and Augenstein, Isabelle},
  title     = {Discourse-aware Rumour stance classification in social media using sequential classifiers},
  journal   = {Inf. Process. Manage.},
  volume    = {54},
  number    = {2},
  pages     = {273--290},
  year      = {2018},
  NOurl     = {https://doi.org/10.1016/j.ipm.2017.11.009},
  NOdoi     = {10.1016/j.ipm.2017.11.009}
}

@InProceedings{mikolov-yih-zweig:2013:NAACL-HLT,
  author    = {Mikolov, Tomas  and  Yih, Wen-tau  and  Zweig, Geoffrey},
  title     = {Linguistic Regularities in Continuous Space Word Representations},
  booktitle = {Proceedings of the 2013 Conference of the North American Chapter of the Association for Computational Linguistics: Human Language Technologies},
  series    = {NAACL-HLT~'13},
  NOmonth   = {June},
  year      = {2013},
  address   = {Atlanta, GA, USA},
  NOpublisher = {Association for Computational Linguistics},
  pages     = {746--751},
  NOurl       = {http://www.aclweb.org/anthology/N13-1090}
}

@InProceedings{rosenthal-farra-nakov:2017:SemEval,
  author    = {Rosenthal, Sara  and  Farra, Noura  and  Nakov, Preslav},
  title     = {{SemEval}-2017 Task 4: Sentiment Analysis in {T}witter},
  booktitle = {Proceedings of the 11th International Workshop on Semantic Evaluation},
  series = {SemEval~'17},
  NOmonth     = {August},
  year      = {2017},
  address   = {Vancouver, Canada},
  NOpublisher = {Association for Computational Linguistics},
  pages     = {502--518},
  NOurl       = {http://www.aclweb.org/anthology/S17-2088}
}

@article{Baeza-Yates:2018:BW:3229066.3209581,
 author = {Baeza-Yates, Ricardo},
 title = {Bias on the Web},
 journal = {Commun. ACM},
 issue_date = {June 2018},
 volume = {61},
 number = {6},
 month = may,
 year = {2018},
 issn = {0001-0782},
 pages = {54--61},
 numpages = {8},
 NOurl = {http://doi.acm.org/10.1145/3209581},
 doi = {10.1145/3209581},
 acmid = {3209581},
 NOpublisher = {ACM},
 NOaddress = {New York, NY, USA},
}

@inproceedings{balikas2017multitask,
  title={Multitask Learning for Fine-Grained {T}witter Sentiment Analysis},
  author={Balikas, Georgios and Moura, Simon and Amini, Massih-Reza},
  booktitle={Proceedings of the 40th International ACM SIGIR Conference on Research and Development in Information Retrieval},
  series = {SIGIR~'17},
  pages={1005--1008},
  year={2017},
  address = {Tokyo, Japan},
  NOorganization={ACM}
}

@inproceedings{yu2006collaborative,
  title={Collaborative ordinal regression},
  author={Yu, Shipeng and Yu, Kai and Tresp, Volker and Kriegel, Hans-Peter},
  booktitle={Proceedings of the 23rd international conference on Machine learning},
  pages={1089--1096},
  year={2006},
  organization={ACM}
}


@inproceedings{esuli2016isti,
  title={ISTI-CNR at SemEval-2016 Task 4: Quantification on an ordinal scale},
  author={Esuli, Andrea},
  booktitle={Proceedings of the 10th International Workshop on Semantic Evaluation (SemEval-2016)},
  pages={92--95},
  year={2016}
}

@inproceedings{he2016yzu,
  title={{YZU-NLP} team at SemEval-2016 task 4: Ordinal sentiment classification using a recurrent convolutional network},
  author={He, Yunchao and Yu, Liang-Chih and Yang, Chin-Sheng and Lai, K Robert and Liu, Weiyi},
  booktitle={Proceedings of the 10th International Workshop on Semantic Evaluation},
  series = {SemEval~'16},
  address = {San Diego, CA, USA},
  pages={251--255},
  year={2016}
}

@inproceedings{kulkarni2018multi,
  title       = {Multi-view Models for Political Ideology Detection of News Articles},
  author      = {Kulkarni, Vivek and Ye, Junting and Skiena, Steven and Wang, William Yang},
  booktitle   = {Proceedings of the Conference on Empirical Methods in Natural Language Processing},
  NOmonth     = {October},
  year        = {2018},
  series      = {EMNLP~'18},
  pages = {3518--3527},
  address = {Brussels, Belgium},
  NOpublisher = {Association for Computational Linguistics},
}

@article{10.2307/27798525,
 NOISSN = {00030554, 15375943},
 NOURL = {http://www.jstor.org/stable/27798525},
 author = {Margit Tavits and Natalia Letki},
 journal = {The American Political Science Review},
 number = {4},
 pages = {555--569},
 publisher = {[American Political Science Association, Cambridge University Press]},
 title = {When Left Is Right: Party Ideology and Policy in {Post-Communist Europe}},
 volume = {103},
 year = {2009}
}

@article{Barron:19,
  author = {Barr\'{o}n-Cede\~no, Alberto and
  Da San Martino, Giovanni and
  Jaradat, Israa and
  Nakov, Preslav},
  title = {Proppy: Organizing News Coverage on the Basis of Their Propagandistic Content},
  journal = {Information Processing and Management},
  year = {2019}
}

@InProceedings{AAAI2019:proppy,
author = {Barr\'{o}n-Cede\~no, Alberto and
Da San Martino, Giovanni and
Jaradat, Israa and
Nakov, Preslav},
title = {Proppy: A System to Unmask Propaganda in Online News},
booktitle = {Proceedings of the Thirty-Third AAAI Conference on Artificial Intelligence},
series = {AAAI'19},
year = {2019},
address = {Honolulu, HI, USA},
NOmonth = {January},
}

@article{JDIQ2019,
    author = {Preslav Nakov and M\`{a}rquez, Llu\'{i}­s and Barr\'{o}n-Cede\~{n}o, Alberto and Pepa Gencheva and Georgi Karadzhov and Tsvetomila Mihaylova and Mitra Mohtarami and James Glass},
    title = {Automatic Fact Checking Using Context and Discourse Information},
    journal = {ACM Journal of Data and Information Quality (ACM JDIQ)},
    year = {2019},
}

@inproceedings{P14-1105,
    title = "Political Ideology Detection Using Recursive Neural Networks",
    author = "Iyyer, Mohit  and
      Enns, Peter  and
      Boyd-Graber, Jordan  and
      Resnik, Philip",
    booktitle = "Proceedings of the 52nd Annual Meeting of the Association for Computational Linguistics",
    seris = {ACL~'14},
    NOmonth = jun,
    year = "2014",
    address = "Baltimore, MD, USA",
    NOpublisher = "Association for Computational Linguistics",
    NOurl = "https://www.aclweb.org/anthology/P14-1105",
    NOdoi = "10.3115/v1/P14-1105",
    pages = "1113--1122",
}

@inproceedings{Gerrish:2011:PLR:3104482.3104544,
 author = {Gerrish, Sean M. and Blei, David M.},
 title = {Predicting Legislative Roll Calls from Text},
 booktitle = {Proceedings of the 28th International Conference on International Conference on Machine Learning},
 series = {ICML~'11},
 year = {2011},
 NOisbn = {978-1-4503-0619-5},
 address = {Bellevue, Washington, USA},
 pages = {489--496},
 numpages = {8},
 NOurl = {http://dl.acm.org/citation.cfm?id=3104482.3104544},
 NOacmid = {3104544},
 NOpublisher = {Omnipress},
 NOaddress = {USA},
} 

@inproceedings{D13-1010,
    title = "Measuring Ideological Proportions in Political Speeches",
    author = "Sim, Yanchuan  and
      Acree, Brice D. L.  and
      Gross, Justin H.  and
      Smith, Noah A.",
    booktitle = "Proceedings of the 2013 Conference on Empirical Methods in Natural Language Processing",
    series = {EMNLP~'13},
    NOmonth = oct,
    year = "2013",
    address = "Seattle, WA, USA",
    NOpublisher = "Association for Computational Linguistics",
    NOurl = "https://www.aclweb.org/anthology/D13-1010",
    pages = "91--101",
}

@InProceedings{P17-1068,
  author = 	"Preo{\c{t}}iuc-Pietro, Daniel
		and Liu, Ye
		and Hopkins, Daniel
		and Ungar, Lyle",
  title = 	"Beyond Binary Labels: Political Ideology Prediction of {T}witter Users",
  booktitle = 	"Proceedings of the 55th Annual Meeting of the Association for      Computational Linguistics (Volume 1: Long Papers)    ",
  year = 	"2017",
  NOpublisher = 	"Association for Computational Linguistics",
  series = {ACL~'17},
  pages = 	"729--740",
  address = 	"Vancouver, Canada",
  NOdoi = 	"10.18653/v1/P17-1068",
  NOurl = 	"http://aclweb.org/anthology/P17-1068"
}
\bibliographystyle{acl_natbib}

\end{document}